\documentclass[a4,11pt]{article}
\usepackage{graphicx} 
\usepackage{float}
\usepackage{amsmath,amsthm}
\usepackage{amsfonts}
\usepackage{latexsym}
\usepackage{amssymb}
\usepackage{setspace}
\usepackage{comment}
\usepackage{ascmac}
\usepackage{cases}
\usepackage{color}

\makeatletter
 
  \@addtoreset{equation}{section}
 \makeatother
\begin{document}
\title{{Mean-field equilibrium price formation with exponential utility}
\footnote{
Forthcoming in {\it Stochastics and Dynamics}. 
All the contents expressed in this research are solely those of the authors and do not represent any views or 
opinions of any institutions. The authors are not responsible or liable in any manner for any losses and/or damages caused by the use of any contents in this research.
}
}

\author{Masaaki Fujii\footnote{mfujii@e.u-tokyo.ac.jp, https://orcid.org/0000-0002-3332-068X, Graduate School of Economics, The University of Tokyo, Tokyo, Japan }, \quad
Masashi Sekine \footnote{sekinemasashi@g.ecc.u-tokyo.ac.jp, 
https://orcid.org/0009-0002-5042-9895,  Graduate School of Economics, The University of Tokyo, Tokyo, Japan }
}
\date{ 
The 1st version: October 10, 2023\\
This version: January 6, 2025 
}
\maketitle



\newtheorem{definition}{Definition}[section]
\newtheorem{assumption}{Assumption}[section]
\newtheorem{condition}{$[$ C}
\newtheorem{lemma}{Lemma}[section]
\newtheorem{proposition}{Proposition}[section]
\newtheorem{theorem}{Theorem}[section]
\newtheorem{remark}{Remark}[section]
\newtheorem{example}{Example}[section]
\newtheorem{corollary}{Corollary}[section]
%
\def\cala{{\cal A}}
\def\calb{{\cal B}}
\def\calc{{\cal C}}
\def\cald{{\cal D}}
\def\cale{{\cal E}}
\def\calf{{\cal F}}
\def\calg{{\cal G}}
\def\calh{{\cal H}}
\def\cali{{\cal I}}
\def\calj{{\cal J}}
\def\calk{{\cal K}}
\def\call{{\cal L}}
\def\calm{{\cal M}}
\def\caln{{\cal N}}
\def\calo{{\cal O}}
\def\calp{{\cal P}}
\def\calq{{\cal Q}}
\def\calr{{\cal R}}
\def\cals{{\cal S}}
\def\calt{{\cal T}}
\def\calu{{\cal U}}
\def\calv{{\cal V}}
\def\calw{{\cal W}}
\def\calx{{\cal X}}
\def\caly{{\cal Y}}
\def\calz{{\cal Z}}
%
\def\sskip{\hspace{0.5cm}}
\def\simleq{ \raisebox{-.7ex}{\em $\stackrel{{\textstyle <}}{\sim}$} }
\def\leqsim{ \raisebox{-.7ex}{\em $\stackrel{{\textstyle <}}{\sim}$} }
\def\nn{\nonumber}
\def\be{\begin{equation}}
\def\ee{\end{equation}}
\def\bea{\begin{eqnarray}}
\def\eea{\end{eqnarray}}
%

\def\calf{{\cal F}}
\def\wt{\widetilde}
\def\mbb{\mathbb}
\def\ol{\overline}
\def\ul{\underline}
\def\sign{{\rm{sign}}}
\def\wh{\widehat}
\def\mg{\mathfrak}
\def\display{\displaystyle}

\def\vr{\varrho}
\def\ep{\epsilon}

\def\Prb{\mbb{P}}
\def\del{\delta}
\def\Del{\Delta}

\def\deln{\delta_{\mathfrak{n}}}
\def\oldeln{\overline{\delta}_{\mathfrak{n}}}
\def\vep{\varepsilon}

\def\red{\textcolor{red}}
\def\Ito{{It\^o}~}
\def\blangle{\bigl\langle}
\def\Blangle{\Bigl\langle}
\def\brangle{\bigr\rangle}
\def\Brangle{\Bigr\rangle}
\def\bi{\begin{itemize}}
\def\ei{\end{itemize}}
\def\ac{\acute}
\def\pr{\prime}
\def\mgn{\mathfrak{n}}

\def\part{\partial}
\def\ul{\underline}
\def\ol{\overline}
\def\vp{\varpi}
\def\nn{\nonumber}
\def\be{\begin{equation}}
\def\ee{\end{equation}}
\def\bea{\begin{eqnarray}}
\def\eea{\end{eqnarray}}
\def\bg{\boldsymbol}
\def\bull{$\bullet~$}
\def\ex{\mbb{E}}
\def\opb{\wh{\beta}}
\def\zo{{0,1}}
\def\gmone{{\gamma^1}}
\vspace{0mm}

\begin{abstract}
In this paper, using the mean-field game theory, we study a problem of equilibrium price formation among many investors with exponential utility 
in the presence of liabilities unspanned by the security prices. 
The investors are heterogeneous in their initial wealth, risk-averseness parameter, as well as stochastic liability at the terminal time.
We characterize the equilibrium risk-premium process of the risky stocks in terms of the solution to 
a novel mean-field backward stochastic differential equation (BSDE),  whose driver has quadratic growth both in the stochastic integrands 
and in their conditional expectations. We prove the existence of a solution to the mean-field BSDE under several conditions
and show that the resultant risk-premium process actually clears the market in the large population limit.
\end{abstract}

\section{Introduction}
The key objective of theories of equilibrium price formation is to find asset prices for which the 
financial market is in equilibrium.  Equilibrium prices are prices for which the optimal decisions of all the agents, given these asset prices, 
are such that the aggregate demand of each asset equals its aggregate supply.
Constructing and investigating the properties of the equilibrium prices are one of the major issues in financial economics.
See, for example,  Back~\cite{Back} and Munk~\cite{Munk} for details on this topic. 
The existence of  market equilibrium in complete markets
is well known, the details of which can be found in Karatzas \& Shreve~\cite{Karatzas-Shreve}.
However, the situation in incomplete markets has been unclear except for
finite dimensional economic models, such as the one in Cuoco \& He~\cite{Cuoco-1}.
Although a set of sufficient conditions for the existence of a representative agent has been derived, 
confirming these conditions for given models involves difficult mathematical problems and still remains open.
See discussions in Jarrow~\cite[Section 14.5]{Jarrow-2} and references therein. 
In particular, it has been known that obtaining a concrete characterization of 
the equilibrium prices is very difficult when there are idiosyncratic liabilities, labor incomes or endowments that are 
unspanned by the security prices. 
This is a quite unsatisfactory situation, because we can imagine that the existence of variety of idiosyncratic shocks among the agents 
induced from these 
terms is the very reason why we observe lively transactions in financial markets.
In this paper, with the help of the recent developments of the mean-field game (MFG) theory, we propose 
a new concise technique to tackle the problem of equilibrium price formation in 
an incomplete market under an exponential utility.

Since the pioneering works of Lasry \& Lions~\cite{Lions-1, Lions-2, Lions-3} and 
Huang et al.~\cite{Caines-Huang, Caines-Huang-2, Caines-Huang-3}, significant developments in the MFG
theory have made us possible to understand some of the long-standing issues of multi-agent games.
If the interactions among the agents are symmetric, then
the MFG techniques can render, in the large population limit,  a very complex problem of solving a large coupled system of equations
that characterizes a Nash equilibrium feasible by 
transforming it into separate and simpler problems of the optimization for a representative agent
and of finding a fixed point. The resultant solution in the mean field equilibrium is known 
to provide an $\epsilon$-Nash equilibrium for the original game with a large but a finite number of agents.
The details of the MFG theory and many applications can be found, for example, 
in two volumes by Carmona \& Delarue~\cite{Carmona-Delarue-1, Carmona-Delarue-2} and 
in a lecture note by Cardaliaguet~\cite{Cardaliaguet-note}.

We want to construct the price process of risky stocks
(more precisely, the associated risk-premium process) {\it endogenously} 
so that they balance the demand and supply among a large number of financial firms facing the  market-wide common noise as well as
their own idiosyncratic noise.  
Unfortunately, this {\it market clearing condition} does not fit well to the concept of Nash equilibrium.
Actually, if we change a trading strategy of one agent away from her equilibrium solution while keeping the other agents'
strategies unchanged, then the balance of demand and supply will inevitably be broken down.
Since the MFG theory has been developed primarily for the Nash games, 
most of its applications have not treated the market-clearing equilibrium.
In fact, in many of the existing examples,  their primary interests are not in the price formation 
and the asset price process is typically assumed exogenously. 
For example, let us refer to \cite{Espinosa-Touzi, Frei-DosReis, Fu-Su-Zhou, Fu-Zhou, Lacker-Z} 
as related works dealing with optimal investment problems with exponential and power utilities.
All of these works concern with the Nash equilibrium among the investors competing in a relative performance criterion,
while the relevant price processes are given exogenously.

Recently, there also has been progress in the MFG theory for the problem of equilibrium price formation 
under the market clearing condition. 
Gomes \& Sa\'ude~\cite{Gomes-Saude} present a deterministic model of electricity price.
Its extension with random supply is given by Gomes et al.~\cite{Gomes-random-supply}.
The same authors also study, in \cite{Gomes-random-noise}, a price formation problem of a commodity whose production 
is subject to random fluctuations.
Evangelista et al.~\cite{Evangelista-game-liquidity} investigate the price formation of an asset being traded in a limit order book
and show promising numerical results using the actual high-frequency data of the listed stocks in several exchanges.
Shrivats et al.~\cite{Firoozi-1} deal with a price formation problem for the solar renewable energy certificate (SREC)
by solving forward-backward stochastic differential equations (FBSDEs) of McKean-Vlasov type,
and Firoozi et al.~\cite{Firoozi-2} deal with a principal-agent problem in the associated emission market.
Fujii \& Takahashi~\cite{Fujii} solve a stochastic mean-field price model of securities
in the presence of stochastic order flows, and in \cite{Fujii-SC},  the same authors prove the strong convergence to
the mean-field limit from the setup with finite number of agents. 
Fujii \& Takahashi~\cite{Fujii-Major} extend the above model to the presence of a major player.
Recently,  Fujii~\cite{Fujii-CP} develops a model that allows the co-presence of cooperative and non-cooperative populations
to learn how the price process is formed when a group of agents act in a coordinated manner.

But still, there exist two important restrictions  in all of these papers given above:
firstly,  the relevant control of each agent is interpreted as the {\it trading rate}
that is absolutely continuous with respect to the Lebesgue measure $dt$; secondly, the  cost function of each agent consists
of  terms representing some penalties on the trading speed and on the inventory size of the assets.
In other words, from the perspective of financial applications, 
the existing results cannot deal with the general self-financing trading strategies
nor the utility (or cost) functions given directly in terms of the associated wealth process of  the portfolio.\footnote{See also a very recent work by Lavigne \& Tankov~\cite{Lavigne-Tankov}
which adopts a very different approach to investigate the mean-field equilibrium of the carbon emissions among the firms.}
Hence the question of equilibrium price formation, at least in the standard and traditional setup in financial economics, remains unanswered.
Here, a major problem in dealing with a utility function of 
wealth has been the difficulties in guaranteeing the convexity of the Hamiltonian 
associated with the Pontryagin's maximum principle and in obtaining enough regularity to 
prove the well-posedness of the associated FBSDEs.

In this paper, we investigate an equilibrium price formation of the risky stocks to address the above two concerns by using the method of Hu, Imkeller \& M\"uller \cite{Hu-Imkeller} and the mean field game theory.
In contrast to the existing works using the method of \cite{Hu-Imkeller}, such as \cite{Fu-Su-Zhou,Fu-Zhou}, which found an MFG equilibrium with relative performance criterion, we establish a market-clearing equilibrium for the price-formation problem.
Our goal is to construct the risk-premium process endogenously
so that the demand and supply of the associated stocks always balance among a large number of financial firms (agents) who are allowed to deploy
general self-financing trading strategies. 
We assume that the agents have a common type of preference specified by an exponential utility with respect to their wealth.
The agents have heterogeneities in their initial wealth, 
the size of risk-averseness parameter as well as the stochastic liability at the terminal time.
We allow the liabilities (not restricted to be positive) to depend both on the common noise
and on the idiosyncratic noise so that
the they can describe idiosyncratic shocks of each agent in addition to the market-wide financial shocks.
For example, since our agents are financial firms, it is realistic to assume that they have significant derivative liabilities.
Although payoffs specified by financial contracts usually depend only on the common  information, the volume of the contracts
can naturally depend also on idiosyncratic information of each agent,  such as her cooperate size, cooperate culture, and her client base. 
Personnel expenses of each agent can also be included in her liability.

For solving the optimization problem of each agent, we adopt the optimal martingale principle developed by Hu, Imkeller \& M\"uller \cite{Hu-Imkeller},
instead of the Pontryagin's maximum principle. Their method can be applied to many popular classes of utility functions, 
such as Exponential, Power and Log types. 
Recently, Xing~\cite{Hao-Xing} solved the optimization problem for  Epstein-Zin recursive utility by the same method.
In every case, the relevant equation characterizing the optimality is given by a quadratic-growth BSDE (qg-BSDE)~\cite{Kobylanski}.
In contrast to the existing literature, 
we cannot simply assume the risk-premium to be a bounded process,  
since this condition may be too stringent to achieve a market-clearing equilibrium.
In fact, we will see that it is necessary to relax the boundedness assumption
on the risk-premium process in general. 
Since the standard results on a qg-BSDE cannot handle the existence of such a process in its driver, we need a special treatment to 
show its well-posedness. Fortunately enough, the special structure of the BSDE inherited from the exponential utility  allows us to solve the problem
by a standard application of the comparison principle.

After we obtain the optimal strategy of each agent, we move on to characterize the mean-field equilibrium 
in terms of  a novel mean-field BSDE whose driver has quadratic growth  both in the 
stochastic integrands and in their conditional expectations. The new BSDE is interesting in its own right since 
the same type of equations may become relevant for similar applications of mean-field equilibrium to other utility functions. 
As main contributions of this work, we show the existence of a solution to this novel mean-field BSDE under several conditions, and then prove that the risk-premium process expressed by its solution actually clears the market in the large population limit.

The organization of the paper is as follows. After explaining some notations in 
Section~\ref{sec-notation}, we solve the optimization problem of each agent with exponential utility in Section~\ref{sec-each-optimization}.
In particular, a special attention is paid to allow the unbounded risk-premium process.
In Section~\ref{sec-mean-field-BSDE}, we derive a novel mean-field BSDE and prove its existence of a solution 
under several conditions. We show that the risk-premium process characterized by the solution of this BSDE
actually clears the market in the large population limit in Section~\ref{sec-market-clearing}.
We concludes the paper by Section~\ref{sec-conclusion}.

\section{Notations}
\label{sec-notation}
Throughout this work $T>0$ denotes a given time horizon.
For random variables and stochastic processes defined on a given complete probability space $(\Omega,\calf, \mbb{P})$
endowed with a filtration $\mbb{F}:=(\calf_t)_{t\in[0,T]}$ satisfying the usual conditions, 
we use the following conventions to represent frequently used function spaces:
\bi
\item $\mbb{H}^2(\mbb{P}, \mbb{F};\mbb{R}^{1\times d})$ (or simply $\mbb{H}^2$) denotes the set of $\mbb{R}^{1\times d}$-valued $\mbb{F}$-progressively
measurable processes $Z$ satisfying
\be
\|Z\|_{\mbb{H}^2}:=\mbb{E}^{\mbb{P}}\Bigl[\int_0^T |Z_t|^2 dt\Bigr]^\frac{1}{2}<\infty. \nn
\ee
\item $\mbb{H}^2_{\rm BMO}(\mbb{P}, \mbb{F};\mbb{R}^{1\times d})$ (or simply $\mbb{H}^2_{\rm BMO}$)
is a subset of $\mbb{H}^2(\mbb{P}, \mbb{F};\mbb{R}^{1\times d})$ satisfying
\be
\|Z\|^2_{\mbb{H}^2_{\rm BMO}}:=\sup_{\tau\in \calt}\Bigr\|\mbb{E}^{\mbb{P}}\Bigl[\int_\tau^T |Z_t|^2 dt|\calf_\tau\Bigr]\Bigr\|_{\infty}
<\infty. 
\label{eq-h2bmo-norm}
\ee
Here, $\calt$ is the set of $\mbb{F}$-stopping times with values in $[0,T]$, and $\|\cdot\|_{\infty}$ denotes the $\mbb{P}$-essential 
supremum over $\Omega$.
In this case, thanks to the result of Kazamaki~\cite{Kazamaki},
the Dolean-Dade exponential $\cale\Bigl(\int_0^\cdot Z_s dW_s\Bigr)_t,~t\in[0,T]$ is known to be uniformly integrable (i.e.~of class D), and 
$\Bigl(\int_0^\cdot Z_s dW_s\Bigr)_{t\in[0,T]}$ is called a BMO-martingale. Here, $W$ is a $d$-dimensional standard Brownian motion defined on $(\Omega, \calf, \mbb{P};\mbb{F})$. 
As an important property of $Z\in \mbb{H}^2_{\rm BMO}$, let us mention the so-called energy inequality: For every $n\in \mbb{N}$, 
the next inequality holds;
\be
\mbb{E}^{\mbb{P}}\Bigl[\Bigl(\int_0^T |Z_s|^2 ds\Bigr)^n\Bigr]\leq n!\bigl(\|Z\|^2_{\mbb{H}^2_{\rm BMO}}\bigr)^n.
\label{ineq-energy}
\ee
See \cite{Cvitanic-Zhang}[Lemma 9.6.5]. This space is used as a class of the risk-premium process.
Remark~\ref{mu-theta} gives some details on this choice.
\item $\mbb{S}^2(\mbb{P},\mbb{F};\mbb{R}^d)$ (or simply $\mbb{S}^2$)
is the set of $\mbb{R}^d$-valued $\mbb{F}$-adapted continuous processes $X$ satisfying
\be
\|X\|_{\mbb{S}^2}:=\mbb{E}^{\mbb{P}}\bigl[\sup_{t\in[0,T]}|X_t|^2\bigr]^\frac{1}{2}<\infty. \nn
\ee
\item $\mbb{S}^\infty(\mbb{P},\mbb{F};\mbb{R}^d)$ (or simply $\mbb{S}^\infty$)
is a subset of $\mbb{S}^2(\mbb{P},\mbb{F};\mbb{R}^d)$ satisfying
\be
\|X\|_{\mbb{S}^\infty}:=\bigl\|\sup_{t\in[0,T]} |X_t|\bigr\|_{\infty}<\infty. \nn
\ee 
\ei
For an $m\times n$ matrix $A$, we denote its kernel and range by ${\rm Ker}(A):=\{x\in \mbb{R}^n; Ax=0\}$ 
and ${\rm Range}(A):=\{y\in \mbb{R}^m; y=Ax, x\in \mbb{R}^n\}$, respectively.
When there is no risk of confusion, the specification such as $(\mbb{P},\mbb{F},\mbb{R}^d)$ or part of it will be often omitted for notational simplicity.
\\

The relevant probability spaces used in the first part of this work are given below.
\bi
\item $(\Omega^0, \calf^0, \mbb{P}^0)$ is a complete probability space with a complete and right-continuous filtration 
$\mbb{F}^0:=(\calf_t^0)_{t\in[0,T]}$ generated by $d_0$-dimensional standard Brownian motion $W^0:=(W_t^0)_{t \in[0,T]}$.
We set $\calf^0:=\calf^0_T$.
This space is used to model market-wide noise and information common to all the agents.
\item $(\Omega^1, \calf^1, \mbb{P}^1)$ is a complete probability space with a complete and right-continuous filtration
$\mbb{F}^1:=(\calf_t^1)_{t\in [0,T]}$ generated by $d$-dimensional standard Brownian motion $W^1:=(W^1_t)_{t\in [0,T]}$
and a $\sigma$-algebra $\sigma(\xi^1,\gamma^1)$, which defines $\calf_0^1$, generated by a bounded $\mbb{R}$-valued 
random variable $\xi^1$ and a strictly positive bounded random variable $\gamma^1$. We set $\calf^1:=\calf^1_T$.
This space is used to model idiosyncratic noise and information for an agent (agent-1).
In later sections, we create independent copies $(\Omega^i,\calf^i, \mbb{P}^i)_{i\in \mbb{N}}$ of this  space endowed with 
$\mbb{F}^i:=(\calf_t^i)_{t \in[0,T]}$ to model idiosyncratic information for a large number of agents, (agent-$i$, $i\in\mbb{N}$).
\item $(\Omega^\zo, \calf^\zo, \mbb{P}^\zo)$ is a probability space defined on the product set $\Omega^\zo:=\Omega^0\times \Omega^1$
with $(\calf^\zo,\mbb{P}^\zo)$ the completion of $(\calf^0\otimes \calf^1, \mbb{P}^0\otimes \mbb{P}^1)$.
$\mbb{F}^{0,1}:=(\calf^{0,1}_t)_{t\in [0,T]}$ denotes the complete and right-continuous augmentation of $(\calf_t^0\otimes \calf_t^1)_{t\in[0,T]}$.
A generic element of $\Omega^\zo$
is denoted by $\omega:=(\omega^0,\omega^1)\in \Omega^0\times \Omega^1$.
\item $\calt^\zo$ is the set of all $\mbb{F}^{0,1}$-stopping times with values in $[0,T]$.
\item $\calt^0$ is the set of all $\mbb{F}^0$-stopping times with values in $[0,T]$.
\ei

Throughout this work, we do not distinguish a random variable  defined on a marginal probability space
with its trivial extension to a product space for notational simplicity.
For example, we will use a same symbol $X$ for a random variable $X(\omega^0)$ defined on 
the space $(\Omega^0,\calf^0,\mbb{P}^0)$  and for its trivial extension $X(\omega^0,\omega^1):=X(\omega^0)$
defined on  the product space $(\Omega^\zo, \calf^\zo,\mbb{P}^\zo)$.

\section{Exponential utility optimization for a given agent}
\label{sec-each-optimization}
In this section, we consider the optimization problem for an agent (agent-1) whose preference is given by an exponential 
utility. We characterize her optimal trading strategy in terms of the quadratic-growth BSDE 
by the approach proposed by Hu, Imkeller \& M\"uller~\cite{Hu-Imkeller}.
In particular, however, in order to deal with the mean-field price formation as in \cite{Fujii},
we need to relax their assumption on the boundedness of the risk-premium process  $\theta:=(\theta_t)_{t\in[0,T]}$ to the unbounded one in $\mbb{H}^2_{\rm BMO}$.

\subsection{The market and utility function}
The market dynamics and the (agent-1)'s idiosyncratic environment are modeled on the filtered probability space $(\Omega^\zo,\calf^\zo,
\mbb{P}^\zo;\mbb{F}^\zo)$ defined in the previous section. In this section, 
the expectation with respect to $\mbb{P}^\zo$ is simply denoted by $\mbb{E}[\cdot]$. The financial market is specified as follows.

\begin{assumption}
\label{assumption-market}
{\rm (i)} The risk-free interest rate is  zero.\\
{\rm (ii)} There are $n\in \mbb{N}$ non-dividend paying risky stocks whose price dynamics is given by
\be
		S_t=S_0+\int_0^t {\rm diag}(S_s)(\mu_s ds+\sigma_s dW_s^0), ~t\in[0,T] 
\label{eq-stock-price}
\ee
where $S_0:=(S_0^i)_{i=1}^n, S_0^i>0$ is the $n$-vector denoting the initial stock prices, $\mu:=(\mu_t)_{t\in[0,T]}$ is an $\mbb{R}^n$-valued, $\mbb{F}^0$-progressively
measurable process belonging to $\mbb{H}^2_{\rm BMO}(\mbb{P}^{0,1},\mbb{F}^0)$~\footnote{Clearly, $\mu$ 
is also in $\mbb{H}^2_{\rm BMO}(\mbb{P}^\zo,\mbb{F}^\zo)$. In fact,  additional information from $\mbb{F}^1$
cannot increase the $\mbb{H}^2_{\rm BMO}$-norm $(\ref{eq-h2bmo-norm})$ 
since $\mu$ is $\mbb{F}^0$-adapted which is independent from $\mbb{F}^1$.}.
$\sigma:=(\sigma_t)_{t\in[0,T]}$ is an $\mbb{R}^{n\times d_0}$-valued, bounded, and $\mbb{F}^0$-progressively measurable process
 such that  there exist positive constants $0<\ul{\lambda}<\ol{\lambda}$ satisfying 
\be
\ul{\lambda}I_n\leq (\sigma_t\sigma_t^\top)\leq \ol{\lambda}I_n, \quad {\text{$dt\otimes \mbb{P}^0$-a.e.}} \nn
\ee 
Here, $I_n$ denotes $n\times n$ identity matrix.
\end{assumption}

\begin{remark}
    $\mu\in \mbb{H}^2_{\rm BMO}$ is a very strong assumption. 
    In fact, typical Gaussian processes such as the Ornstein-Uhlenbeck process are not contained in this class. 
    This is unfortunate, however, such an assumption is essential to deal with the quadratic-growth BSDE, which will be introduced later.
\end{remark}

Since the interest rate is zero, the risk-premium process  $\theta:=(\theta_t)_{t\in[0,T]}$ is defined by $\theta_t:=\sigma_t^\top
(\sigma_t\sigma_t^\top)^{-1}\mu_t$. Hence, for any $t\geq 0$, $\theta_t\in {\rm Range}(\sigma_t^\top)={\rm Ker}(\sigma_t)^\perp$.
Here, $\top$ denotes the transposition and ${\rm Ker} (\sigma_t)^\perp$ the orthogonal complement of ${\rm Ker}(\sigma_t)$ in $\mbb{R}^{d_0}$.
Note that $\theta$ is in  $\mbb{H}^2_{\rm BMO}$ due to the boundedness of the process $\sigma$. 
By the regularity of $(\sigma\sigma^\top)$, we have ${\rm rank}(\sigma)=n$. 
The financial market is incomplete even within the common information
$\mbb{F}^0$ in general, since we have $n\leq d_0$. Recall that we have additional noise associated with $\mbb{F}^1$ in the liability given below.

\begin{remark}
We emphasize that, although $\mu$ is unbounded, the dynamics of stock price $(\ref{eq-stock-price})$ 
is well defined. The easiest way to check this is to change the probability measure to the
risk-neutral one, which is possible since $\theta\in \mbb{H}^2_{\rm BMO}$ (See, Kazamaki~\cite{Kazamaki-T}). Under this equivalent probability measure,
the stock price process is a uniquely specified as a martingale.
\end{remark}

\begin{definition}
For each $s\in[0,T]$, let us denote by
\be
L_s:=\{u^\top \sigma_s; u\in \mbb{R}^n\} \nn
\ee
the  linear subspace of $\mbb{R}^{1\times d_0}$ spanned by the $n$ row vectors of $\sigma_s$.
For any $z\in \mbb{R}^{1\times d_0}$,  $\Pi_s(z)$ denotes the orthogonal projection of $z$ onto the 
linear subspace $L_s$.
\end{definition}
\noindent 
Notice that $\theta_s^\top\in L_s$ for every $s\in[0,T]$ by its construction.

\begin{remark}
The projections to the range as well as the kernel of the finite dimensional matrix are Borel measurable.
It then follows that the process $(\Pi_s(z_s))_{s\in[0,T]}$ is progressively measurable if  so is the process $(\sigma_s, z_s)_{s\in[0,T]}$.
See \cite[Chapter 1; Lemma 4.4, Corollary 4.5]{Karatzas-Shreve} for the proof.
\end{remark}

The idiosyncratic environment for the agent-1 is modeled by a triple $(\xi^1, \gamma^1, F^1)$.
\begin{assumption}
\label{assumption-agent}
{\rm (i)} $\xi^1$ is an $\mbb{R}$-valued, bounded, and $\calf_0^1$-measurable random variable denoting the initial wealth for the agent-1.\\
{\rm (ii)} $\gamma^1$ is an $\mbb{R}$-valued, bounded, and $\calf_0^1$-measurable random variable, satisfying
\be
\ul{\gamma}\leq \gamma^1\leq \ol{\gamma}, \nn
\ee
with some positive constants $0<\ul{\gamma}\leq \ol{\gamma}$. $\gamma^1$ denotes the size of risk-averseness of the agent-1. \\
{\rm (iii)} $F^1$ is an $\mbb{R}$-valued, bounded, and $\calf^\zo_T$-measurable random variable denoting the liability of the agent-1 at time $T$. \\
{\rm (iv)} The agent-1 has a negligible market share and hence her trading activities have no impact on the stock prices, i.e., 
she is a price taker. 
\end{assumption}
\begin{remark}
Notice that the liability $F^1$ is subject to common shocks from $\calf_T^0$ as well as idiosyncratic shocks from $\calf_T^1$.
\end{remark}

The wealth process of the agent-$1$ under the self-financing trading strategy $\pi$ is given by
\be
\calw_t^{1,\pi}=\xi^1+\sum_{j=1}^n\int_0^t\frac{\pi_{j,s}}{S^j_s}dS^j_s=\xi^1+\int_0^t \pi_s^\top \sigma_s(dW_s^0+\theta_s ds). \nn
\ee
Here, $\pi:=(\pi_t)_{t\in[0,T]}$ is an $\mbb{R}^n$-valued, $\mbb{F}^\zo$-progressively measurable process
representing the invested amount of money in each of the $n$ stocks.
The problem of the agent-$1$ is to solve
\be
\sup_{\pi \in \mbb{A}^1} U^1(\pi), \nn
\ee
where the functional $U^1$ is called exponential utility (for the agent-$1$). It is defined by
\be
U^1(\pi):=\mbb{E}\Bigl[-\exp\Bigl(-\gamma^1 \Bigl(\xi^1+\int_0^T \pi_s^\top \sigma_s (dW_s^0+\theta_s ds)-F^1\Bigr)\Bigr)\Bigr]. 
\label{eq-utility-functional}
\ee
It means that the low performance in the sense of $\calw^{1,\pi}_T-F^1<0$ is punished heavily and
the high performance $\calw^{1,\pi}_T-F^1>0$ is only weakly valued. 

In this paper, we will not delve into the concrete modeling of the liability, which can include common as well as idiosyncratic shocks associated with,
for example, financial market,  endowment, consumption, local price of commodities,  and/or budgetary target imposed on the agent by her manager.
\begin{definition}
\label{def-admissible-space}
The admissible space $\mbb{A}^1$ is the set of all $\mbb{R}^n$-valued, $\mbb{F}^\zo$-progressively measurable trading strategies $\pi$
that satisfy
$
\mbb{E}\Bigl[ \int_0^T |\pi_s^\top \sigma_s|^2ds \Bigr]<\infty, \nn
$
and such that
\be
\bigl\{\exp(-\gamma^1 \calw_\tau^{1,\pi}); \tau \in \calt^\zo\bigr\} \nn
\ee
is uniformly integrable (i.e. of class D). We also define
$
\cala^1:=\bigl\{p=\pi^\top \sigma; \pi\in \mbb{A}^1\bigr\}. \nn
$
\end{definition}

\noindent
Note that $p$ is an $\mbb{R}^{1\times d_0}$-valued process with $p_s\in L_s$ for any $s\in[0,T]$.
The problem for the agent-1 can be equivalently said to find the value function:
\be
V^{1,*}:=\sup_{p\in \cala^1}\mbb{E}\Bigl[-\exp\Bigl(-\gamma^1\Bigl(\xi^1+\int_0^T p_s (dW_s^0+\theta_sds)-F^1\Bigr)\Bigr)\Bigr]. \nn
\ee

\begin{remark}
\label{mu-theta} Before closing this section, let us give some remarks on the choice of $\mbb{H}^2_{\rm BMO}$ 
for the class of $\theta$. In this work, our goal is  to find an appropriate $\theta$ (and hence $\mu$)
to achieve an equilibrium. If we put an agent in a stochastic environment as we did, her demand/supply of goods becomes inevitably stochastic. 
In order to take balance among many agents, the risk-premium process $\theta$ must also  be stochastic in general.
This is already the case in the classical results for the complete market with finite number of agents (See Chapter 4 in Karatzas~\&~Shreve~\cite{Karatzas-Shreve}).
In order not to miss a valuable candidate of equilibrium, 
it is of course desirable to weaken assumptions on $\theta $ as much as possible,  but we need mathematical tractability at the same time.
We have chosen the class of $\mbb{H}^2_{\rm BMO}$ particularly because that it makes Dol\'eans-Dade exponential uniformly integrable for {\it every finite time interval} $[0,T]$, which is necessary to justify the measure change we need. As we will see below, it also turns out to be the class of 
the stochastic integrands $Z$ of qg-BSDEs with solutions in $\mbb{S}^\infty$.  This plays a crucial role to 
maintain the consistency of our logic. See Sections~\ref{sec-mean-field-BSDE} and \ref{sec-market-clearing}.
\end{remark}

\subsection{ Characterization of the optimal trading strategy}
Thanks to the work~\cite{Hu-Imkeller}, we can characterize the optimal trading strategy by using a solution to a certain BSDE (instead of FBSDEs)
in a rather straightforward way. When a utility (or equivalently cost) function has a special homothetic form as in the current case,
their method often provides much simpler description of the optimal strategy than in the case where the  Pontryagin's maximum principle is applied.

We try to construct a family of stochastic processes $\{R^p:=(R^p_t)_{t\in[0,T]}, p\in \cala^1\}$ satisfying the following properties:
\begin{definition} 
(Condition-R) \\
{\rm (i)} $R_T^p=-\exp\bigl(-\gamma^1 (\calw_T^{1,p}-F^1)\bigr)$ a.s. for all $p\in\cala^1$.\\
{\rm (ii)} $R_0^p=R_0$ a.s. for all $p\in \cala^1$ and for some $\calf_0^1~(=\calf^\zo_0)$-measurable random variable $R_0$.\\
{\rm (iii)} $R^p$ is an $(\mbb{F}^\zo,\mbb{P}^\zo)$-supermartingale for all $p\in \cala^1$, and there exists some $p^*\in \cala^1$
such that $R^{p^*}$ is an $(\mbb{F}^\zo,\mbb{P}^\zo)$-martingale.
\end{definition}
\noindent
In fact, if we can find such a family $\{R^p\}$, then for any $p\in \cala^1$, we have
\be
\mbb{E}\bigl[-\exp\bigl(-\gamma^1(\calw_T^{1,p}-F^1)\bigr)\bigr]\leq \mbb{E}[R_0]=\mbb{E}\bigl[-\exp\bigl(-\gamma^1(\calw_T^{1,p^*}-F^1)\bigr)\bigr], \nn
\ee
and hence $p^*$ is an optimal trading strategy for the agent-1.

In order to construct the family $\{R^p\}$, we try to find an appropriate process $Y:=(Y_t)_{t\in[0,T]}$ with
which the process $R^p$ is given by 
\be
R_t^p=-\exp\bigl(-\gamma^1(\calw_t^{1,p}-Y_t)\bigr), \quad t\in[0,T], ~p\in \cala^1.
\label{eq-R-hypo}
\ee 
Here, the triple $(Y,Z^0,Z^1)$, which  is a $(\mbb{R},\mbb{R}^{1\times d_0}, \mbb{R}^{1\times d})$-valued process, 
is an $\mbb{F}^\zo$-adapted solution to the BSDE
\be
Y_t=F^1+\int_t^T f(s,Z_s^0,Z_s^1)ds-\int_t^T Z_s^0 dW_s^0-\int_t^T Z_s^1 dW_s^1, \quad t\in[0,T]. \nn
\ee
The concrete form of  the driver $f$ is to be determined below so that $\{R^p\}$ satisfies the desired properties.  

Under the hypothesis of $(\ref{eq-R-hypo})$, we get, by \Ito formula,
\be
\begin{split}
dR_t^p&=R_t^p\Bigl(-\gmone d(\calw_t^{1,p}-Y_t)+\frac{(\gamma^1)^2}{2}d\langle \calw^{1,p}-Y\rangle_t\Bigr)\\
&=R_t^p\Bigl(-\gamma^1 (p_t\theta_t+f(t,Z_t^0,Z_t^1))+\frac{(\gamma^1)^2}{2}(|p_t-Z_t^0|^2+|Z_t^1|^2)\Bigr)dt\\
&\quad+R_t^p\bigl(-\gamma^1(p_t-Z_t^0)dW_t^0+\gamma^1 Z_t^1 dW_t^1\bigr), \quad t\in[0,T]. \nn
\end{split}
\ee 
In order to guess an appropriate form of $f$, let us formally solve it as
\be
\begin{split}
R_t^p&=-\exp\bigl(-\gmone(\xi^1-Y_0)\bigr)\exp\Bigl(\int_0^t\bigl[-\gmone(p_s\theta_s+f(s,Z_s^0,Z_s^1))+\frac{(\gmone)^2}{2}(|p_s-Z_s^0|^2+|Z_s^1|^2)\bigr]ds\Bigr)\\
&\quad \times \cale\Bigl(\int_0^\cdot\bigl[-\gmone(p_s-Z_s^0)dW_s^0+\gmone Z_s^1 dW_s^1\bigr]\Bigr)_t. \nn
\end{split}
\ee
We want to set the driver $f(s,Z_s^0, Z_s^1)$ so that, for all $s\in[0,T]$, 
\bi
\item $-\gmone(p_s\theta_s+f(s,Z_s^0,Z_s^1))+\frac{(\gmone)^2}{2}(|p_s-Z_s^0|^2+|Z_s^1|^2)\geq 0$ for all $p\in \cala^1$,
\item $\exists p^*\in \cala^1$ such that $-\gmone(p_s^*\theta_s+f(s,Z_s^0,Z_s^1))+\frac{(\gmone)^2}{2}(|p_s^*-Z_s^0|^2+|Z_s^1|^2)=0$.
\ei

The above conditions suggest that
\be
\begin{split}
f(s,Z_s^0,Z_s^1)&=\inf_{p_s\in L_s}\Bigl\{-p_s\theta_s+\frac{\gmone}{2}(|p_s-Z_s^0|^2+|Z_s^1|^2)\Bigr\}\\
&=\inf_{p_s\in L_s}\Bigl\{ \frac{\gamma^1}{2}\Bigl|p_s-\Bigl(Z_s^0+\frac{\theta_s^\top}{\gamma^1}\Bigr)\Bigr|^2-Z_s^0\theta_s-\frac{1}{2\gamma^1}|\theta_s|^2
+\frac{\gamma^1}{2}|Z_s^1|^2\Bigr\}. \nn
\end{split}
\ee
This is a special case treated by \cite{Hu-Imkeller}[Section 2] 
with a trading constraint $\pi_t\in \wt{C}$ by a general closed subset $\wt{C}\subset \mbb{R}^{n}$,
which is now replaced by the whole space $\mbb{R}^{n}$. A candidate of the optimal strategy $p^*$ is then given by
\be
p^*_t=Z_t^{0\parallel}+\frac{\theta_t^\top}{\gmone}, ~t\in[0,T].
\label{optimal-p}
\ee
Here, for notational simplicity, we have written $Z_s^{0\parallel}:=\Pi_s(Z_s^0)$ and $Z_s^{0\perp}:=Z_s^0-\Pi_s(Z_s^0)$.
They are orthogonal each other and $|Z_s^0|^2=|Z_s^{0\parallel}|^2+|Z_s^{0\perp}|^2$. Recall that $\Pi_s(\theta_s^\top)=\theta_s^\top$ for every $s$.
With this convention, we have
\be
\begin{split}
f(s,Z_s^0,Z_s^1)&=-Z_s^0\theta_s-\frac{1}{2\gamma^1}|\theta_s|^2+\frac{\gamma^1}{2}(|Z_s^{0\perp}|^2+|Z_s^1|^2 )\\
&=-Z_s^{0\parallel}\theta_s-\frac{1}{2\gamma^1}|\theta_s|^2+\frac{\gamma^1}{2}(|Z_s^{0\perp}|^2+|Z_s^1|^2 )\\
\end{split}
\ee
where we used the fact $\theta_s^\top \in L_s$ and hence $Z_s^{0\perp}\theta_s=0$ in the second equality.

Therefore, the associated qg-BSDE is given by
\be
Y_t=F^1+\int_t^T \Bigl(-Z_s^{0\parallel}\theta_s-\frac{|\theta_s|^2}{2\gmone}+\frac{\gmone}{2}(|Z_s^{0\perp}|^2+|Z_s^1|^2)\Bigr)ds-\int_t^T Z_s^0 dW_s^0-\int_t^T Z_s^1 dW_s^1,
~t\in[0,T]. 
\label{BSDE-org}
\ee
In order to make its appearance simpler, we rewrite the equation with $G^1:=\gmone F^1$, and $(y, z^0,z^1):=(\gmone Y,\gmone Z^0,\gmone Z^1)$.
Then, we can equivalently work on the normalized BSDE,
\be
y_t=G^1+\int_t^T\Bigl(-z_s^{0\parallel}\theta_s-\frac{1}{2}|\theta_s|^2+\frac{1}{2}(|z_s^{0\perp}|^2+|z_s^1|^2)\Bigr)ds-\int_t^T z_s^0 dW_s^0-\int_t^T z_s^1 dW_s^1,~t\in[0,T].
\label{BSDE-norm}
\ee
In this case, $p^*$ is given by $\gamma^1 p^*_t=z_t^{0\parallel}+\theta_t^\top$. There should be no confusion which BSDE is being discussed by checking the terminal function and the presence of $\gamma^1$.
In order to conclude that they are actually  what we want, we need to verify that the resultant family $\{R^p\}$ $(\ref{eq-R-hypo})$ and the process $p^*$ $(\ref{optimal-p})$ satisfy (Condition-R).

\subsection{Solution of the BSDE and its verification}

We  emphasize that, in contrast to the work~\cite{Hu-Imkeller}, our risk-premium process $\theta\in \mbb{H}^2_{\rm BMO}$ is unbounded in general.
As we will see in later sections, this generalization is necessary to handle the mean-field market clearing equilibrium.
Due to this unbounded risk-premium process,  we cannot apply the  standard results on qg-BSDEs given by Kobylanski~\cite{Kobylanski}. 
Moreover, since the exponential integrability of $(|\theta_t|^2, t\in[0,T])$
is not guaranteed in general, we cannot apply the extensions on the qg-BSDE theories such as \cite{Briand-Hu, Briand-Hu-2, Hu-Tang}, either.
In particular, the case $(|\theta_t|^{1+\alpha}, \alpha<1)$ is covered by the result in $\cite{Hu-Tang}$ but not the case where 
$|\theta_t|^2$-term is contained in the driver. 
Fortunately, thanks to the special form of its driver inherited from the exponential utility, 
we can show the existence of a unique solution $(y,z^0,z^1)$ to the BSDE $(\ref{BSDE-norm})$ (and equivalently $(\ref{BSDE-org})$) 
in the space $\mbb{S}^\infty\times \mbb{H}^2_{\rm BMO}\times 
\mbb{H}^2_{\rm BMO}$ by a simple modification of the standard approach \cite{Kobylanski}.

\begin{lemma}
\label{lemma-Z-norm}
Let Assumptions~\ref{assumption-market} and \ref{assumption-agent} be in force.
If there exists a bounded (with respect to the $y$-component) solution, i.e. $(y,z^0,z^1)\in \mbb{S}^\infty\times \mbb{H}^2 \times \mbb{H}^2$,
to the BSDE $(\ref{BSDE-norm})$, then $z:=(z^0,z^1)$ is in $\mbb{H}^2_{\rm BMO}$.  
\begin{proof}
This can be proved by a simple application of It\^o formula. We give the details in Appendix~\ref{sec-A-1}.
\end{proof}
\end{lemma}

The above lemma is now used to guarantee the uniqueness of solution if $y \in \mbb{S}^\infty$.
\begin{theorem}
\label{th-uniqueness}
Let Assumptions~\ref{assumption-market} and \ref{assumption-agent} be in force.
If the solution to $(\ref{BSDE-norm})$ is bounded, i.e. $(y,z^0,z^1)\in \mbb{S}^\infty\times \mbb{H}^2\times \mbb{H}^2$, then 
such a solution is unique.
\begin{proof}
Suppose that there are two bounded solutions $(y,z^0,z^1)$ and $(\ac{y},\ac{z}^0,\ac{z}^1)$.
By Lemma~\ref{lemma-Z-norm}, we know  that $(z^0,z^1)$ and $(\ac{z}^0, \ac{z}^1)$ are actually in $\mbb{H}^2_{\rm BMO}$.
Let us put;
\be
\Del y_t:=y_t-\ac{y}_t, \quad \Del z_t^0:=z_t^0-\ac{z}_t^0, \quad \Del z_t^1:=z_t^1-\ac{z}_t^1. \nn
\ee
From the orthogonality between $z^{0\parallel}$ and $z^{0\perp}$, we have
\be
\begin{split}
\Del y_t&=\int_t^T\Bigl(-\Del z_s^{0\parallel} \theta_s+\frac{1}{2}\Del (z_s^{0\perp})(z_s^{0\perp}+\ac{z}_s^{0\perp})^\top+
\frac{1}{2}\Del z_s^1(z_s^1+\ac{z}_s^1)^\top\Bigr)ds-\int_t^T \Del z_s^0 dW_s^0-\int_t^T \Del z_s^1 dW_s^1\\
&=\int_t^T\Bigl(-\Del z_s^0 \Bigl(\theta_s-\frac{1}{2}(z_s^{0\perp}+\ac{z}_s^{0\perp})^\top\Bigr)+
\frac{1}{2}\Del z_s^1(z_s^1+\ac{z}_s^1)^\top\Bigr)ds-\int_t^T \Del z_s^0 dW_s^0-\int_t^T \Del z_s^1 dW_s^1\\
&=-\int_t^T \Del z_s^0\Bigl(dW_s^0+\bigl(\theta_s-\frac{1}{2}(z_s^{0\perp}+\ac{z}_s^{0\perp})^\top\bigr)ds\Bigr)
-\int_t^T \Del z_s^1\Bigl(dW_s^1-\frac{1}{2}(z_s^1+\ac{z}_s^1)^\top ds\Bigr)\\
&=-\int_t^T \Del z_s^0d\wt{W}_s^0
-\int_t^T \Del z_s^1d\wt{W}_s^1,  \nn
\end{split}
\ee
where we have defined a new measure $\wt{\mbb{P}}$ equivalent to $\mbb{P}^\zo$ by
\be
\frac{d\wt{\mbb{P}}}{d\mbb{P}^\zo}\Bigr|_{\calf^\zo_t}:=M_t:=\cale\Bigl(-\int_0^\cdot \bigl(\theta_s^\top-
\frac{1}{2}(z_s^{0\perp}+\ac{z}_s^{0\perp})\bigr) dW_s^0+\int_0^\cdot \frac{1}{2}(z_s^1+\ac{z}_s^1)dW_s^1\Bigr)_t,~t\in[0,T]
\nn
\ee
and 
$
(\wt{W}_t^0, \wt{W}_t^1)_{t\in[0,T]}
$
denote the standard Brownian motions under $\wt{\mbb{P}}$. 
This measure change is well-defined since $(\theta^\top, z^{0\perp}+\ac{z}^{0\perp}, z^1+\ac{z}^1)$ are in $\mbb{H}^2_{\rm BMO}$ and hence $M$ is a uniformly integrable martingale.
By the result of Kazamaki~\cite{Kazamaki-T} and \cite{Kazamaki}[Remark 3.1], the following so-called reverse H\"older inequality holds:
\be
\mbb{E}\bigl[M_T^r|\calf_t^\zo\bigr]\leq C M_t^r, \nn
\ee
where $C>0$ and $r>1$ are some constants depending only on the $\mbb{H}^2_{\rm BMO}$-norm of $(\theta^\top, z^{0\perp}+\ac{z}^{0\perp}, z^1+\ac{z}^1)$. 
With $q=\frac{r}{r-1}>1$ and $j=0,1$, H\"older and the energy inequality $(\ref{ineq-energy})$ imply
\be
\begin{split}
\mbb{E}^{\wt{\mbb{P}}}\Bigl[\int_0^T |\Del z_s^j|^2ds\Bigr]&=\mbb{E}\Bigl[M_T\Bigl(\int_0^T |\Del z_s^j|^2 ds\Bigr)\Bigr]\\
&\leq \mbb{E}[M_T^r]^\frac{1}{r}\mbb{E}\Bigl[\Bigl(\int_0^T |\Del z_s^j|^2 ds\Bigr)^q\Bigr]^\frac{1}{q}<\infty. \nn
\end{split}
\ee
Thus $\Del y$ is an $(\mbb{F}^{0,1},\wt{\mbb{P}})$-martingale.
Thus we can conclude  that $\Del y=0$ and so are $(\Del z^0, \Del z^1)$.
\end{proof}
\end{theorem}

Since $\theta$ is in $\mbb{H}^2_{\rm BMO}$, it is natural to change the measure to absorb the term $(-z^{0\parallel}\theta)~(=-z^0\theta)$ in the driver of $(\ref{BSDE-norm})$.
Let us define the measure $\mbb{Q}~(\sim \mbb{P}^\zo)$ by
\be
\frac{d\mbb{Q}}{d\mbb{P}^\zo}\Bigr|_{\calf^\zo_t}:=\cale\Bigl(-\int_0^\cdot \theta_s^\top dW_s^0\Bigr)_t, \nn
\ee
where the standard Brownian motions under $\mbb{Q}$ are given by
\be
\wt{W}_t^0=W_t^0+\int_0^t \theta_s ds, \quad \wt{W}_t^1=W_t^1, \quad t\in[0,T]. \nn
\ee
Therefore, instead of $(\ref{BSDE-norm})$, we can equivalently work on the BSDE defined on $(\Omega^\zo,\calf^\zo,\mbb{Q};\mbb{F}^\zo)$
endowed with the Brownian motions $(\wt{W}^0, \wt{W}^1)$;
\be
y_t=G^1+\int_t^T \Bigl(-\frac{1}{2}|\theta_s|^2+\frac{1}{2}(|z_s^{0\perp}|^2+|z_s^1|^2)\Bigr)ds-\int_t^T z_s^0d\wt{W}^0_s-\int_t^T z_s^1d\wt{W}_s^{1},
~t\in[0,T].
\label{BSDE-norm-Q}
\ee
Although in general, the filtration $\mbb{F}^\zo$ is bigger than the one generated by $(\wt{W}^0,\wt{W}^1)$,
we can still apply the standard techniques of BSDEs. This is  due to 
the stability property of the martingale representation under the absolutely continuous 
measure changes. See \cite{HWY}[Theorem 13.12] for general case and \cite{Jeanblanc}[Section 1.7.7] for Brownian case.
Moreover, by Kazamaki~\cite{Kazamaki}[Theorem 3.3], $\theta$ is still in $\mbb{H}^2_{\rm BMO}(\mbb{Q},\mbb{F}^\zo)$.
Obviously, BSDE $(\ref{BSDE-norm})$ (and equivalently $(\ref{BSDE-org})$) has a bounded solution if and only if  BSDE $(\ref{BSDE-norm-Q})$ has a bounded solution.  

\begin{theorem}
\label{th-qgBSDE-existence}
Let Assumptions~\ref{assumption-market} and \ref{assumption-agent} be in force.
Then there is a unique bounded solution $(y,z^0,z^1)\in \mbb{S}^\infty(\mbb{Q},\mbb{F}^\zo)\times \mbb{H}^2_{\rm BMO}(\mbb{Q},\mbb{F}^\zo)\times \mbb{H}^2_{\rm BMO}(\mbb{Q},\mbb{F}^\zo)$ to the BSDE $(\ref{BSDE-norm-Q})$.
\begin{proof}
Since we work under the measure $\mbb{Q}$ throughout this proof,  we write $\mbb{E}[\cdot]$ instead of $\mbb{E}^{\mbb{Q}}[\cdot]$
for notational simplicity.
Firstly, for each $n\in \mbb{N}$, we consider the truncated BSDE;
\be
y_t^n=G^1+\int_t^T \Bigl(-\frac{1}{2}(|\theta_s|^2\wedge n)+\frac{1}{2}(|z_s^{n,0\perp}|^2+|z_s^{n,1}|^2)\Bigr)ds-\int_t^T z_s^{n,0}d\wt{W}_s^0
-\int_t^T z_s^{n,1}d\wt{W}_s^1, \quad t\in[0,T].  
\label{BSDE-truncated}
\ee
Clearly, the truncated BSDE $(\ref{BSDE-truncated})$ has a 
unique bounded solution $(y^n,z^{n,0},z^{n,1})\in \mbb{S}^\infty\times \mbb{H}^2_{\rm BMO}\times \mbb{H}^2_{\rm BMO}$ by 
the standard result of \cite{Kobylanski}. 
Moreover, by the comparison principle~\footnote{The comparison principle tells that the size of the solution $y$
responds monotonically with respect to that of the terminal and driver functions.
See, for example, \cite{Zhang-book}[Theorem~7.3.1].}
obtained in the same work, we have $y^{n+1}\leq y^n$ for all 
$n\in \mbb{N}$~\footnote{Here, we use  $|z_s^{n,0\perp}|^2-|z_s^{n+1,0\perp}|^2=\Del (z_s^{n,0\perp})(z_s^{n,0\perp}+z_s^{n+1,0\perp})^\top
=\Del z_s^{n,0}(z_s^{n,0\perp}+z_s^{n+1,0\perp})^\top$ 
to absorb it into the stochastic integral.}.
In particular, uniformly in $n\in \mbb{N}$, the solution $y^n$ is bounded from above as $y^n\leq \ol{y}$, where $\ol{y}$ is the solution to 
another quadratic-growth BSDE;
\be
\ol{y}_t=G^1+\int_t^T \frac{1}{2}\bigl(|\ol{z}_s^{0}|^2+|\ol{z}_s^1|^2\bigr)ds-\int_t^T \ol{z}_s^0d\wt{W}_s^0-\int_t^T \ol{z}_s^1 d\wt{W}_s^1, \quad t\in[0,T]. \nn
\ee
It also has a unique bounded solution $(\ol{y},\ol{z}^0,\ol{z}^1)\in \mbb{S}^\infty\times \mbb{H}^2_{\rm BMO}\times \mbb{H}^2_{\rm BMO}$ 
with $\|\ol{y}\|_{\mbb{S}^\infty}\leq \|G^1\|_{\infty}$  by the standard result.

Once again, by the comparison principle, $y^n$  is also bounded from below uniformly in $n\in\mbb{N}$ as $\ul{y}\leq y^n$,
where $\ul{y}$ is the solution to the next simple BSDE;
\be
\ul{y}_t=G^1+\int_t^T \Bigl(-\frac{1}{2}|\theta_s|^2 \Bigr)ds-\int_t^T \ul{z}_s^0 d\wt{W}_s^0-\int_t^T \ul{z}_s^1 d\wt{W}_s^1, \quad t\in[0,T]. \nn
\ee
Obviously, it has a unique solution $(\ul{y}, \ul{z}^0,\ul{z}^1)\in \mbb{S}^2\times \mbb{H}^2\times \mbb{H}^2$.
Moreover, for any $t\in[0,T]$, 
\be
\begin{split}
\ul{y}_t&=\mbb{E}[G^1|\calf^\zo_t]-\frac{1}{2}\mbb{E}\Bigl[\int_t^T |\theta_s|^2 ds|\calf^\zo_t\Bigr] \\
&\geq -\Bigl(\|G^1\|_{\infty}+\frac{1}{2}\|\theta\|^2_{\mbb{H}^2_{\rm BMO}}\Bigr)>-\infty. \nn
\end{split}
\ee
Hence we conclude that, uniformly in $n\in \mbb{N}$, $y^n$ satisfies the following bound,
\be
-\Bigl(\|G^1\|_{\infty}+\frac{1}{2}\|\theta\|^2_{\mbb{H}^2_{\rm BMO}}\Bigr)\leq y^n\leq \|G^1\|_{\infty}. 
\label{yn-bound}
\ee

Since $\{y^n\}$ is bounded from below and it is monotonically decreasing in $n\in \mbb{N}$, we can define 
the process $y:=(y_t)_{t\in[0,T]}$ by
\be
y=\lim_{n\rightarrow \infty }y^n. \nn
\ee
Moreover, by repeating the proof of Lemma~\ref{lemma-Z-norm}, we get from the estimate $(\ref{yn-bound})$
\be
\forall n\in \mbb{N}, \quad \|(z^{n,0},z^{n,1})\|_{\mbb{H}^2_{\rm BMO}}^2\leq \exp\bigl(4\|G^1\|_{\infty}+2\|\theta\|^2_{\mbb{H}^2_{\rm BMO}}\bigr).  \nn
\ee
In particular, $(z^{n,0}, z^{n,1})_{n\in \mbb{N}}$ are weakly relatively compact in $\mbb{H}^2$ and hence, under an appropriate subsequence (still denoted by $n$), we have $\exists (z^0,z^1)\in \mbb{H}^2\times \mbb{H}^2$, such that
\be
z^{n,0}\rightharpoonup z^0, \quad z^{n,1}\rightharpoonup z^1\quad {\text{ weakly in $\mbb{H}^2$ as $n\rightarrow \infty$}}. \nn
\ee
The remaining procedures to show the triple $(y,z^0,z^1)$ actually solves $(\ref{BSDE-norm-Q})$ are the same as those in \cite{Kobylanski}.
See also \cite{Cvitanic-Zhang}[Section 9.6]. For readers convenience, we shall give the details in Appendix~\ref{sec-A-2}.
\end{proof}
\end{theorem}

\begin{corollary}
\label{corollary-existence}
Let Assumptions~\ref{assumption-market} and \ref{assumption-agent} be in force.
Then the BSDE $(\ref{BSDE-org})$ $(\text{resp.} ~(\ref{BSDE-norm}) )$ has a
unique bounded solution $(Y,Z^0,Z^1)$ $(\text{resp.}~ (y,z^0,z^1))$ in $\mbb{S}^\infty(\mbb{P}^\zo,\mbb{F}^\zo)\times \mbb{H}^2_{\rm BMO}(\mbb{P}^\zo,\mbb{F}^\zo)\times  \mbb{H}^2_{\rm BMO}(\mbb{P}^\zo,\mbb{F}^\zo)$.
\end{corollary}

We are now ready to verify the Condition-R.
\begin{theorem}
Let Assumptions~\ref{assumption-market} and \ref{assumption-agent} be in force. Then
the family of processes $\{R^p, p\in \cala^1 \}$ defined by $(\ref{eq-R-hypo})$ with the process $Y$ as the unique bounded solution to
the BSDE $(\ref{BSDE-org})$ satisfies the Condition-R, and the process $p^*$ given by $(\ref{optimal-p})$
gives the unique $($up to $dt\otimes \mbb{P}^\zo$-null set$)$ optimal trading strategy for the agent-1.
\begin{proof}
From $(\ref{eq-R-hypo})$, we have
\be
R_t^p=-\exp\bigl(-\gamma^1(\calw_t^{1,p}-Y_t)\bigr)=-\exp\bigl(-\gmone\calw_t^{1,p}+y_t\bigr), ~t\in[0,T], \nn
\ee
and 
\be
R_0^p=-\exp(-\gamma^1 \xi^1+y_0) \nn
\ee 
for all $p\in\cala^1$. 
Here, $y:=(y_t)_{t\in[0,T]}$ is the rescaled solution of $(\ref{BSDE-norm})$.
Since $y\in \mbb{S}^\infty$, $(R_t^p, t\in[0,T])$ is clearly of class D for any $p\in\cala^1$  by the definition of admissibility $\cala^1$.

Let us choose $p=p^*$ as in $(\ref{optimal-p})$. Then we have
\be
\begin{split}
dR_t^{p^*}&=R_t^{p^*}\bigl(-\gamma^1(p_t^*-Z_t^0)dW_t^0+\gamma^1 Z_t^1 dW_t^1\bigr)\\
&=R_t^{p^*}\bigl(-(\theta_t^\top-z_t^{0\perp}) dW_t^0+z_t^1 dW_t^1\bigr), ~t\in[0,T], \nn
\end{split}
\ee
and hence, for any $t\in[0,T]$, 
\be
\begin{split}
R_t^{p*}&=-\exp\bigl(-\gamma^1 \calw_t^{1,p^*}+y_t\bigr)\\
&=-\exp\bigl(-\gamma^1 \xi^1+y_0\bigr)\cale\Bigl(-\int_0^\cdot (\theta_s^\top-z_s^{0\perp}) dW_s^0+\int_0^\cdot z_s^1 dW_s^1\Bigr)_t. \nn
\end{split}
\ee
Since $(\theta^\top-z^{0\perp}, z^1)$ are in $\mbb{H}^2_{\rm BMO}$ and $(\gamma^1, \xi^1, y_0)$ are all bounded, $R^{p*}$ is a uniformly integrable martingale.
Uniform integrability of $R^{p*}$ and the boundedness of $y$ then imply that $(\exp(-\gamma^1 \calw_t^{1,p^*}))_{t\in[0,T]}$
is also uniformly integrable. Therefore, we obtain the admissibility $p^*\in \cala^1$.
The uniqueness of $p^*$ follows from the strict convexity of 
$-\gamma^1 (p \theta+f(s,z^0,z^1))+\frac{(\gmone)^2}{2}(|p-z^0|^2+|z^1|^2)$
with respect to $p$, which induces a strictly negative drift for $R^p$ if $p\neq p^*$.
Since we know $R^p$ is of class D, its supermartingale property is now obvious.
\end{proof}
\end{theorem}

\begin{remark}
It is important to note that the optimal trading strategy $\pi^*$ (or equivalently $p^*$) is independent from the initial wealth $\xi^1$.
This is a well-known characteristic of  exponential-type utilities.
It follows, combined with the specification of $U^1(\pi)$ in $(\ref{eq-utility-functional})$, that the problem for the agent-1 and her optimal trading
strategy are invariant under the following transformation:
\be
\begin{split}
\xi^1\longrightarrow &~(\xi^1-\ex[F^1|\calf_0^1]), \\
F^1\longrightarrow &~(F^1-\ex[F^1|\calf_0^1]). 
\end{split}
\label{duality-relation}
\ee
\end{remark}

\begin{remark}
The optimization method of \cite{Hu-Imkeller} we used in this section can be applied exactly in the same way even if the market is complete.
However, we are going to find an market-clearing equilibrium based on MFG technique, and it crucially relies on the presence of idiosyncratic 
noise of each agent to simplify the system in the large population limit.
Since this inevitably makes the market incomplete, our strategy in the following sections does not work in a complete market.
We refer to Karatzas~\&~Shreve~\cite{Karatzas-Shreve}[Chapter 4] for the construction of an equilibrium in a complete market,
which uses convex duality and Lagrange multipliers.
\end{remark}

\section{Mean-field equilibrium model}
\label{sec-mean-field-BSDE}

We are now going to investigate a financial market being participated by many agents,
who are interacting each other through the price process of risky stocks.
Recall that, our final goal of this paper is to find an  risk-premium process $\theta=(\theta_t)_{t\geq 0}$ 
of the $n$ risky stocks {\it endogenously} by imposing {\it the market-clearing condition}, which requires 
the demand and supply of the risky stocks to be always balanced among the agents.
In this section, we shall propose a novel  mean-field BSDE with a quadratic-growth driver,
which is expected to provide, at least intuitively,  the characterization of the desired equilibrium in the large population limit.

\subsection{Heuristic derivation of the mean-field BSDE}
Suppose that there are $N\in \mbb{N}$ agents (agent-$i$, $1\leq i\leq N$) participating in
the same financial market given in Assumption~\ref{assumption-market}.
For each $1\leq i\leq N$, the information set of agent-$i$ is provided by the probability space $(\Omega^{0,i},\calf^{0,i}, \mbb{P}^{0,i})$
which is a completion of the product space $(\Omega^0,\calf^0, \mbb{P}^0)\otimes (\Omega^i,\calf^i,\mbb{P}^i)$.
The associated filtration $\mbb{F}^{0,i}:=(\calf_t^{0,i})_{t\in [0,T]}$ is the complete and right-continuous augmentation of 
$(\calf_t^0\otimes \calf_t^i)_{t\in[0,T]}$. $\calt^{0,i}$ is the set of $\mbb{F}^{0,i}$-measurable stopping times with values in $[0,T]$.
Here,  for each $i$, the filtered probability space $(\Omega^i, \calf^i, \mbb{P}^i;\mbb{F}^i)$
is an independent copy of $(\Omega^1,\calf^1,\mbb{P}^1;\mbb{F}^1)$
constructed exactly in the same way  as in Section~\ref{sec-notation} 
with $(\xi^i,\gamma^i, W^i)$ instead of $(\xi^1, \gamma^1, W^1)$. 
Let us define a complete probability space $(\Omega,\calf,\mbb{P})$ with filtration $\mbb{F}:=(\calf_t)_{t\in[0,T]}$ in such a way that
it allows us to model all the agents in a common  space:
\bi
\item $\Omega:=\Omega^0\times \prod_{i=1}^N \Omega^i$
and $(\calf,\mbb{P})$ is the completion of $\bigl(\calf^0\otimes \calf^1\otimes\cdots\otimes\calf^N,
\mbb{P}^0\otimes \mbb{P}^1\otimes \cdots\otimes \mbb{P}^N\bigr)$.
$\mbb{F}$ denotes the complete and the right-continuous augmentation of $(\calf^0_t\otimes \calf_t^1\otimes\cdots\otimes\calf_t^N)_{t\in [0,T]}$.
$\mbb{E}[\cdot]$ denotes the expectation with respect to $\mbb{P}$. 
\ei

We also introduce the liability $F^i$ of the agent-$i$, $1\leq i\leq N$.
Each agent-$i$ is assumed to face the optimization problem 
\be
\sup_{\pi\in \mbb{A}^i}U^i(\pi), \nn
\ee
where the utility functional $U^i$ is defined by 
$
U^i(\pi):=\mbb{E}\Bigl[-\exp\Bigl(-\gamma^i \bigl(\calw_T^{i,\pi}-F^i\bigr)\Bigr)\Bigr], \nn
$
with 
$
\calw_t^{i,\pi}:=\xi^i+\int_0^t \pi_s^\top \sigma_s(dW_s^0+\theta_s ds). \nn
$
The admissible space $\mbb{A}^i$ (and $\cala^i)$ is  defined by the same way as in Definition~\ref{def-admissible-space}
with all the superscripts $``1"$ replaced by $``i"$.
\begin{definition}
The admissible space $\mbb{A}^i$ is the set of all $\mbb{R}^n$-valued, $\mbb{F}^{0,i}$-progressively measurable trading strategies $\pi$
that satisfy
$
\mbb{E}\Bigl[ \int_0^T |\pi_s^\top \sigma_s|^2ds \Bigr]<\infty, \nn
$
and such that
\be
\bigl\{\exp(-\gamma^i \calw_\tau^{i,\pi}); \tau \in \calt^{0,i}\bigr\} \nn
\ee
is uniformly integrable (i.e. of class D). We also define
$
\cala^i:=\bigl\{p=\pi^\top \sigma; \pi\in \mbb{A}^i\bigr\}. \nn
$
\end{definition}

We work under the following assumption.

\begin{assumption}
\label{assumption-hetero}
{\rm (i)} The statements in Assumption~\ref{assumption-agent} hold with $``1"$ replaced by $``i"$, $1\leq i\leq N$.\\
{\rm (ii)} $\{(\xi^i, \gamma^i), 1\leq i\leq N\}$ have the same distribution. In other words, they are independently and identically 
distributed (i.i.d.) on $(\Omega,\calf,\mbb{P})$. \\
{\rm (iii)} $\{F^i, 1\leq i\leq N\}$ are $\calf^0$-conditionally i.i.d.
\end{assumption}

We want to find the risk-premium process $\theta$ that clears the market.
Let us first derive, heuristically,  the relevant mean-field BSDE, which then will be shown to characterize the market-clearing
risk-premium process in the large population limit, by the idea proposed by Fujii \& Takahashi~\cite{Fujii}.

Suppose that a risk-premium process $\theta\in \mbb{H}^2_{\rm BMO}(\mbb{P},\mbb{F}^0;\mbb{R}^{d_0})$ is given.
Recall that $\theta$ has values in ${\rm Range}(\sigma^\top)={\rm Ker}(\sigma)^\perp$, i.e. $\theta_s^\top\in L_s$ for every $s\in[0,T]$. 
By repeating the analysis done in Section~\ref{sec-each-optimization}, one can show that
the unique optimal strategy of agent-$i$, $1\leq i\leq N$, is given by
\be
p^{i,*}_t~(=(\pi_t^{i,*})^\top\sigma_t)=Z_t^{i,0\parallel}+\frac{\theta_t^\top}{\gamma^i}, \quad t\in[0,T], 
\label{p-i-optimal}
\ee
where $Z^{i,0}$ is associated to the solution $(Y^i, Z^{i,0},Z^i)$ of the following BSDE:
\be
Y_t^i=F^i+\int_t^T \Bigl(-Z_s^{i,0\parallel}\theta_s-\frac{|\theta_s|^2}{2\gamma^i}+\frac{\gamma^i}{2}(|Z_s^{i,0\perp}|^2+|Z_s^i|^2)\Bigr)ds
-\int_t^T Z_s^{i,0}dW_s^0-\int_t^T Z_s^i dW_s^i, ~t\in[0,T]. 
\label{BSDE-agent-i}
\ee
Under Assumptions~\ref{assumption-market} and \ref{assumption-hetero}, 
we already know from Corollary~\ref{corollary-existence}
that there is a unique bounded solution $(Y^i,Z^{i,0},Z^i)\in \mbb{S}^\infty(\mbb{P},\mbb{F}^{0,i})
\times \mbb{H}^2_{\rm BMO}(\mbb{P},\mbb{F}^{0,i})\times \mbb{H}^2_{\rm BMO}(\mbb{P},\mbb{F}^{0,i})$ to $(\ref{BSDE-agent-i})$, $1\leq i\leq N$.

\begin{definition}
\label{def-market-clearing}
We say that the market-clearing condition is satisfied if
\be
\frac{1}{N}\sum_{i=1}^N \pi_t^{i,*}=0, \quad dt\otimes \mbb{P}-a.e. \nn
\ee
where $\pi^{i,*}$ is the optimal trading strategy of the agent $i$, $1\leq i\leq N$.
\end{definition}

\begin{remark}
The  market clearing condition in Definition~\ref{def-market-clearing} means that the net supply (demand) of securities is zero in the whole period
under the optimal decisions of all the agents.
For finite $N$, the factor $1/N$ in the above definition is irrelevant.
However, in order to  handle the large population limit $N\rightarrow\infty$, we need the factor $1/N$.
We are going to find a risk-premium process so that the excess demand (or supply)  per capita converges to zero in 
the large population limit.
\end{remark}

The market-clearing condition given by Definiton~\ref{def-market-clearing} requires
that the risk-premium process $\theta$ to be
\be
\theta_t=-\Bigl(\frac{1}{N}\sum_{j=1}^N \frac{1}{\gamma^j}\Bigr)^{-1} \frac{1}{N}\sum_{j=1}^N (Z_t^{j,0\parallel})^\top, \quad t\in[0,T]. 
\label{theta-naive}
\ee
Unfortunately, the suggested $\theta$ by $(\ref{theta-naive})$ is inconsistent with 
our information assumption that requires $\theta$ is $\mbb{F}^0$-adapted i.e. being dependent only on the market-wide information.
However, at this moment, let us {\it formally} consider the $N$-coupled system of quadratic-BSDEs obtained from $(\ref{BSDE-agent-i})$
with $\theta$ replaced by the one given in $(\ref{theta-naive})$; $1\leq i\leq N$, 
\be
\begin{split}
Y_t^i&=F^i+\int_t^T \Bigl\{Z_s^{i,0\parallel}\Bigl(\frac{1}{N}\sum_{j=1}^N \frac{1}{\gamma^j}\Bigr)^{-1}\frac{1}{N}\sum_{j=1}^N (Z_s^{j,0\parallel})^\top
-\frac{1}{2\gamma^i}\Bigl(\frac{1}{N}\sum_{j=1}^N \frac{1}{\gamma^j}\Bigr)^{-2}\Bigl|  \frac{1}{N}\sum_{j=1}^N Z_s^{j,0\parallel} \Bigr|^2 \\
&\hspace{20mm}+\frac{\gamma^i}{2}(|Z_s^{i,0\perp}|^2+|Z_s^i|^2)\Bigr\}ds-\int_t^T Z_s^{i,0}dW_s^0-\int_t^T Z_s^i dW_s^i.
\end{split}
\label{qg-BSDE-system}
\ee

In order to make this system well defined, we need at least additional stochastic integral terms with respect to all of the $(W^j)_{j\neq i}$.
This stems from the martingale representation theorem since $\theta$ is only $\mbb{F}$-adapted. 
We also have to change the space of admissible controls of each agent
to make the associated optimization problem meaningful. 
In any case however, since each agent is a price taker, the interaction among them appears only through the risk-premium process
which has a symmetric form. Thus, from Assumption~\ref{assumption-hetero}, if there is a solution 
$\{(Y^i,Z^{i,0},Z^i), 1\leq i\leq N\}$ to the system $(\ref{qg-BSDE-system})$ (after appropriate modifications), then they are expected to be exchangeable.  In particular,  $(Z^{i,0}_t)_{i=1}^N$ (and hence $(Z_t^{i,0\parallel})_{i=1}^N$) would be  exchangeable random variables, i.e. their joint distribution is invariant under the permutation $\sigma(i)$ of their orders.
If this is the case,  De Finetti's theory of exchangeable sequence of random variables would imply that
\be
\lim_{N\rightarrow \infty}\frac{1}{N}\sum_{i=1}^N Z^{i,0\parallel}_t=\mbb{E}\Bigl[Z_t^{1,0\parallel}|\bigcap_{k\geq 1}\sigma\{Z_t^{j,0\parallel}, j\geq k\}\Bigr]\quad {\text{a.s.}}
\label{De-Finetti-eq}
\ee
See, for example, \cite{Carmona-Delarue-2}[Theorem 2.1].  We can also naturally expect that the tail $\sigma$-field
in $(\ref{De-Finetti-eq})$ converges to $\calf^0_t$ since $(\calf_t^i)_{i\geq 1}$ are independent by construction.
In this way, we can expect from $(\ref{theta-naive})$, at least heuristically, that the equilibrium risk-premium process $\theta$ in the large-$N$ limit may be 
given by
\be
\theta_t=-\wh{\gamma}\mbb{E}[Z_t^{1,0\parallel}|\calf_t^0]^\top=-\wh{\gamma}\mbb{E}[Z_t^{1,0\parallel}|\calf^0]^\top,  
\label{eq-theta-new}
\ee
which is $\mbb{F}^0$-adapted as desired, and  $\wh{\gamma}$ is defined by
\be
\wh{\gamma}:=\frac{1}{\mbb{E}[1/\gamma^1]}. \nn
\ee
The replacement by the conditional expectation $(\ref{eq-theta-new})$ makes the system decoupled since there is no 
interactions among $(Z^{0,j\parallel})_{j\in\mbb{N}}$ any more as opposed to  $(\ref{theta-naive})$.
Such a decoupling phenomenon is usually referred to as {\it propagation of chaos}, and it is the 
main driving force of MFG theory to transform a complex coupled problem into a simple decoupled one.

\vspace{5mm}
The above heuristic discussion motivates us to study the following mean-field BSDE defined
on the filtered probability space $(\Omega^{0,1}, \calf^{0,1},\mbb{P}^{0,1};\mbb{F}^{0,1})$ by choosing the agent-1 as the representative;
\be
Y_t=F^1+\int_t^T \Bigl(\wh{\gamma}Z_s^{0\parallel}\ol{\ex}[Z_s^{0\parallel}]^\top-\frac{\wh{\gamma}^2}{2\gamma^1 }|\ol{\ex}[Z_s^{0\parallel}]|^2+\frac{\gamma^1}{2}(|Z_s^{0\perp}|^2+|Z_s^1|^2)
\Bigr)ds-\int_t^T Z_s^{0} dW_s^0-\int_t^T Z_s^1 dW_s^1,~t\in[0,T].
\label{mfg-BSDE-org}
\ee
Here, we have defined, for any $X\in \mbb{H}^2(\mbb{P}^\zo,\mbb{F}^\zo)$, that
\be
\ol{\ex}[X_t](\omega^0):=\begin{cases} 
\ex[X_t|\calf^0](\omega^0)=\mbb{E}^{\mbb{P}^1}[X_t(\omega^0,\cdot)] \quad {\text{if it exits}}\\
 0\hspace{20mm} {\text{otherwise}}
\end{cases}~.
\nn
\ee
Note that we have $\mbb{E}[X_t|\calf_t^0]=\ex[X_t|\calf^0]$ a.s.~for any $\mbb{F}^{0,1}$-adapted process.  This is because that $X_t$ is independent from  $\sigma(\{W_s^0-W_t^0, s\geq t\})$ and thus the additional information $\calf^0=\calf_T^0\supset \calf_t^0$
does not affect the expectation value. 
As in \cite{Carmona-Delarue-2}[Section 4.3.1], we always choose $\mbb{F}^0$-progressively measurable 
modification of $\ol{\ex}[X]$. In the remainder of this section, we show that there is a solution to this mean-field BSDE under some conditions.
In Section~\ref{sec-market-clearing}, we will show that the risk-premium process defined as 
\be
\theta_t^{\rm mfg}:=-\wh{\gamma}\ol{\ex}[Z_t^{0\parallel}]^\top, \quad t\in[0,T]
\label{theta-mfg}
\ee
by the solution of the mean-field BSDE $(\ref{mfg-BSDE-org})$ actually clears the market 
in the large population limit.
\subsection{Existence of a solution to the mean-field BSDE}
We will work on the filtered probability space $(\Omega^\zo, \calf^\zo, \mbb{P}^\zo;\mbb{F}^\zo)$.
For notational ease, we simply write $(F,\gamma)$ instead of $(F^1,\gamma^1)$ and 
investigate the well-posedness of the mean-field BSDE:
\be
Y_t=F+\int_t^T \Bigl(\wh{\gamma}Z_s^{0\parallel}\ol{\ex}[Z_s^{0\parallel}]^\top-\frac{\wh{\gamma}^2}{2\gamma}|\ol{\ex}[Z_s^{0\parallel}]|^2+\frac{\gamma}{2}(|Z_s^{0\perp}|^2+|Z_s^1|^2)
\Bigr)ds-\int_t^T Z_s^0 dW_s^0-\int_t^T Z_s^1 dW_s^1, \quad t\in[0,T]. 
\label{mfg-BSDE}
\ee
Note that, by Assumption~\ref{assumption-agent} (ii), $\wh{\gamma}$ is a strictly positive constant.

\begin{lemma}
\label{lemma-mfg-bmo}
If there exists a bounded solution $(Y,Z^0,Z^1)\in \mbb{S}^\infty\times \mbb{H}^2\times \mbb{H}^2$ to the BSDE $(\ref{mfg-BSDE})$,
then $Z:=(Z^0,Z^1)$ is in $\mbb{H}^2_{\rm BMO}$.
\begin{proof}
The proof is similar to Lemma~\ref{lemma-Z-norm}. We give the details in Appendix~\ref{sec-A-3}.
\end{proof}
\end{lemma}

Proving the well-posedness of the mean-field BSDE $(\ref{mfg-BSDE})$ is  quite difficult.
First of all, due to the presence of conditional expectations, the comparison principle is not available.
This makes  many of the existing techniques for qg-BSDEs such as in \cite{Kobylanski, Briand-Hu, Briand-Hu-2}
inapplicable. 
Due to the conditional McKean-Vlasov nature of the BSDE $(\ref{mfg-BSDE})$, the traditional approach of MFGs
using Shauder's fixed point theorem does not work, either.
Second, we did not succeed to obtain any {\it a priori} bound on $\|Y\|_{\mbb{S}^\infty}$ nor
any stability result such as \cite{Fujii-qgBSDE}[Lemma 3.3]. Thus, the strategy of constructing a compact set for the decoupling functions
of the regularized BSDEs under the Markovian setup as in \cite{Fujii-ABSDE}[Theorem 4.1] cannot be used.

Therefore, in the following, we will try the method proposed by Tevzadze~\cite{Tevzadze}[Proposition 1].
Among the existing literature on qg-BSDEs,  a unique characteristic of his work  is 
to solve a problem by a contraction mapping theorem without relying on the comparison principle. 
This method is also adopted by Hu et.al.~\cite{Hu-ABSDE} to deal with an anticipated BSDE with quadratic growth terms,  
which is related to an optimization problem with delay.
\\

For notational simplicity, let us define the driver $f$ of the mean-field BSDE by,
for any $(z^0,z^1)\in \mbb{H}^2(\mbb{P}^{0,1},\mbb{F}^{0,1};\mbb{R}^{1\times d_0} \times \mbb{R}^{1\times d})$, 
\be
\begin{split}
f(z_s^0,z_s^1)&:=\wh{\gamma}z_s^{0\parallel}\ol{\ex}[z_s^{0\parallel}]^\top-\frac{\wh{\gamma}^2}{2\gamma}|\ol{\ex}[z_s^{0\parallel}]|^2
+\frac{\gamma}{2}(|z_s^{0\perp}|^2+|z_s^1|^2), \quad s\in[0,T]. \nn
\end{split}
\ee
We have, by completing the square,  
\be
\begin{split}
f(z^0,z^1)&=-\Bigl|\frac{\wh{\gamma}}{\sqrt{2\gamma}}\ol{\ex}[z^{0\parallel}]-\frac{\sqrt{\gamma}}{\sqrt{2}}z^{0\parallel}\Bigr|^2
+\frac{\gamma}{2}|z^{0\parallel}|^2+\frac{\gamma}{2}(|z^{0\perp}|^2+|z^1|^2) \\
&=-\Bigl|\frac{\wh{\gamma}}{\sqrt{2\gamma}}\ol{\ex}[z^{0\parallel}]-\frac{\sqrt{\gamma}}{\sqrt{2}}z^{0\parallel}\Bigr|^2
+\frac{\gamma}{2}(|z^{0}|^2+|z^1|^2). \nn
\end{split}
\ee
Hence, for any $(z^0,z^1)$, 
\be
\begin{split}
f^+(z^0,z^1)&\leq \frac{\ol{\gamma}}{2}(|z^0|^2+|z^1|^2), \\
f^-(z^0,z^1)&\leq 0\vee \Bigl(\Bigl|\frac{\wh{\gamma}}{\sqrt{2\gamma}}\ol{\ex}[z^{0\parallel}]-\frac{\sqrt{\gamma}}{\sqrt{2}}z^{0\parallel}\Bigr|^2
-\frac{\gamma}{2}|z^0|^2\Bigr) \leq \frac{\wh{\gamma}^2}{\ul{\gamma}}|\ol{\ex}[z^{0\parallel}]|^2+\frac{\ol{\gamma}}{2}|z^{0\parallel}|^2, \nn
\end{split}
\ee
where $f^+(z^0,z^1):=\max(f(z^0,z^1),0)$ and $f^-(z^0,z^1):=\max(-f(z^0,z^1),0)$.
Thus, in particular, 
\be
|f(z^0,z^1)|\leq \frac{\ol{\gamma}}{2}(|z^0|^2+|z^1|^2)+\frac{\wh{\gamma}^2}{\ul{\gamma}}|\ol{\ex}[z^{0}]|^2. 
\label{f-abs}
\ee
Moreover,  by Assumption~\ref{assumption-agent} (ii), there is a positive constant $C_\gamma$, which depends only on $(\ul{\gamma},
\ol{\gamma},\wh{\gamma})$,  such that, for any $(z^0,z^1), (\ac{z}^0,\ac{z}^1)\in  \mbb{H}^2$, 
\be
\begin{split}
&|f(z^0,z^1)-f(\ac{z}^0,\ac{z}^1)|\\
&\quad \leq C_\gamma \bigl(|z^0|+|\ac{z}^0|+|z^1|+|\ac{z}^1|+|\ol{\ex}[z^{0}]|+|\ol{\ex}[\ac{z}^{0}]|\bigr)\bigl(|z^0-\ac{z}^0|+|z^1-\ac{z}^1|+
|\ol{\ex}[z^{0}-\ac{z}^{0}]|\bigr).
\end{split}
\label{f-spread}
\ee

Let us observe the following simple fact.
\begin{lemma}
\label{lemma-f-abs}
For any input $(z^0,z^1)\in \mbb{H}^2_{\rm BMO}(\mbb{P}^{0,1},\mbb{F}^\zo)$,
the next  inequality holds;
\be
\sup_{\tau\in\calt^\zo}\Bigl\|\mbb{E}\Bigl[\int_\tau^T |f(z^0_s,z_s^1)|ds|\calf_\tau^\zo\Bigr]\Bigr\|_{\infty}
\leq c_\gamma \|(z^0,z^1)\|^2_{\mbb{H}^2_{\rm BMO}}, \nn
\ee
where $c_\gamma$ is a positive constant given by $\displaystyle c_\gamma:=\frac{\ol{\gamma}}{2}+\frac{\wh{\gamma}^2}{\ul{\gamma}}$.
\begin{proof}
For any input $z^0\in \mbb{H}^2_{\rm BMO}$, $(\ol{\ex}[z^0_t], t\in[0,T])$ is an $\mbb{F}^0$-adapted process and hence independent from $\calf^1$.
This implies, with Jensen's inequality, 
\be
\begin{split}
\sup_{\tau\in \calt^\zo}\Bigr\|\mbb{E}\Bigl[\int_\tau^T |\ol{\ex}[z_s^0]|^2ds|\calf_\tau^\zo\Bigr]\Bigr\|_{\infty}
&=\sup_{\tau\in\calt^0}\Bigr\|\mbb{E}\Bigl[\int_\tau^T |\ol{\ex}[z_s^0]|^2ds|\calf_\tau^0\Bigr]\Bigr\|_{\infty}  \leq \sup_{\tau\in\calt^0}\Bigr\|\mbb{E}\Bigl[\int_\tau^T |z_s^0|^2ds|\calf_\tau^0\Bigr]\Bigr\|_{\infty}.  \nn
\end{split}
\ee
Moreover, for any $\tau\in \calt^0 ~(\subset \calt^\zo)$, we have
$
\mbb{E}\bigl[\int_\tau^T |z_s^0|^2ds|\calf_\tau^0\bigr]=\mbb{E}\Bigl[\mbb{E}\bigl[\int_\tau^T |z_s^0|^2ds|\calf_\tau^\zo\bigr]|\calf_\tau^0\Bigr] \nn
$
and hence
\be
\Bigl\|\mbb{E}\Bigl[\int_\tau^T |z_s^0|^2ds|\calf_\tau^0\Bigr]\Bigr\|_{\infty}\leq \Bigl\|\mbb{E}\Bigl[\int_\tau^T |z_s^0|^2ds|\calf_\tau^\zo\Bigr]
\Bigr\|_{\infty}. \nn
\ee
It follows that
\be
\sup_{\tau\in \calt^\zo}\Bigr\|\mbb{E}\Bigl[\int_\tau^T |\ol{\ex}[z_s^0]|^2ds|\calf_\tau^\zo\Bigr]\Bigr\|_{\infty}
\leq \sup_{\tau\in \calt^\zo}\Bigl\|\mbb{E}\Bigl[\int_\tau^T |z_s^0|^2ds|\calf_\tau^\zo\Bigr]
\Bigr\|_{\infty}.
\label{bmo-comparison}
\ee
Now the conclusion immediately follows from $(\ref{f-abs})$.
\end{proof}
\end{lemma}

\vspace{2mm}
We now define the map $\Gamma: \mbb{H}^2_{\rm BMO}(\mbb{P}^{0,1},\mbb{F}^{0,1};\mbb{R}^{1\times d_0}\times \mbb{R}^{1\times d})\ni (z^0,z^1)\mapsto \Gamma(z^0,z^1)=(Z^0,Z^1)\in 
\mbb{H}^2_{\rm BMO}(\mbb{P}^{0,1},\mbb{F}^{0,1};\mbb{R}^{1\times d_0}\times \mbb{R}^{1\times d})$
by
\be
Y_t=F+\int_t^T f(z^0_s,z^1_s)ds-\int_t^T Z_s^0 dW_s^0-\int_t^T Z_s^1 dW_s^1,
\label{map-Gamma}
\ee
where $\Gamma(z^0,z^1)=(Z^0,Z^1)$ is the stochastic integrands associated to the solution of BSDE $(\ref{map-Gamma})$.

\begin{lemma}
Under Assumption~\ref{assumption-agent}, the map $\Gamma$ is well-defined.
\begin{proof}
For any $(z^0,z^1)\in \mbb{H}^2_{\rm BMO}$, the existence of the unique solution  $(Y,Z^0,Z^1)$ to $(\ref{map-Gamma})$ is obvious.
By taking a conditional expectation, we have from Lemma~\ref{lemma-f-abs},
\be
\|Y\|_{\mbb{S}^\infty}\leq \|F\|_{\infty}+c_\gamma\|(z^0,z^1)\|^2_{\mbb{H}^2_{\rm BMO}}<\infty. \nn
\ee
Moreover, by \Ito formula applied to $|Y_t|^2$, we obtain for any $t\in[0,T]$,
\be
\mbb{E}\Bigl[\int_t^T(|Z^0_s|^2+|Z_s^1|^2)ds|\calf_t^{0,1}\Bigr]\leq \|F\|_{\infty}^2+2\|Y\|_{\mbb{S}^\infty}
\mbb{E}\Bigl[\int_t^T |f(z_s^0,z_s^1)|ds|\calf_t^{0,1}\Bigr]. \nn
\ee
Thus, 
$
\|(Z^0,Z^1)\|_{\mbb{H}^2_{\rm BMO}}^2\leq \|F\|_{\infty}^2+2c_\gamma \|Y\|_{\mbb{S}^\infty}\|(z^0,z^1)\|^2_{\mbb{H}^2_{\rm BMO}}<\infty. \nn
$
\end{proof}
\end{lemma}

For each positive constant $R$, let $\calb_R$ be a subset of $\mbb{H}^2_{\rm BMO}$ defined by
\be
\calb_R:=\Bigl\{ (z^0,z^1)\in \mbb{H}^2_{\rm BMO}(\mbb{P}^{0,1},\mbb{F}^{0,1};\mbb{R}^{1\times d_0}\times \mbb{R}^{1\times d})\Bigr| \|(z^0,z^1)\|^2_{\mbb{H}^2_{\rm BMO}}\leq R^2\Bigr\}. \nn
\ee

\begin{proposition}
\label{prop-Gamma-stability}
Let Assumption~\ref{assumption-agent} be in force. If $ \|F\|_{\infty}\leq 1/(4\sqrt{2}c_\gamma)$,
then with $R:=2\|F\|_{\infty}$, the set $\calb_R$ is stable under the map $\Gamma$, i.e. $(z^0,z^1)\in \calb_R$ implies 
$\Gamma(z^0,z^1)\in \calb_{R}$. Moreover, in this case, the $y$-component of the solution to $(\ref{map-Gamma})$ satisfies $\|Y\|_{\mbb{S}^\infty} \leq R$.
\begin{proof}
By \Ito formula, we have for any $t\in[0,T]$,
\be
|Y_t|^2+\int_t^T (|Z_s^0|^2+|Z_s^1|^2)ds=|F|^2+\int_t^T 2Y_s f(z_s^0,z_s^1)ds-\int_t^T 2Y_s (Z_s^0 dW_s^0+Z_s^1 dW_s^1). \nn
\ee
Taking conditional expectations  we obtain
\be
|Y_t|^2+\mbb{E}\Bigl[\int_t^T (|Z_s^0|^2+|Z_s^1|^2)ds|\calf_t^{0,1}\Bigr]\leq \|F\|^2_{\infty}+2\|Y\|_{\mbb{S}^\infty}\mbb{E}\Bigl[\int_t^T |f(z_s^0,z_s^1)|ds|\calf_t^{0,1}\Bigr].
\label{BR-middle}
\ee
From Lemma~\ref{lemma-f-abs}, taking the essential supremum in the both hands yields,
\be
\underset{(t,\omega)\in[0,T]\times \Omega}{{\rm ess}\sup}\Bigl(|Y_t|^2+\mbb{E}\Bigl[\int_t^T (|Z_s^0|^2+|Z_s^1|^2)ds|\calf_t^{0,1}\Bigr]\Bigr)
\leq \|F\|_{\infty}^2+\frac{1}{2}\|Y\|_{\mbb{S}^\infty}^2+2c_\gamma^2\|(z^0,z^1)\|^4_{\mbb{H}^2_{\rm BMO}}. \nn
\ee
Using the above result and an obvious relation
\be
\begin{split}
\frac{1}{2}\bigl(\|Y\|_{\mbb{S}^\infty}^2+\|(Z^0,Z^1)\|_{\mbb{H}^2_{\rm BMO}}^2\bigr)&\leq \max\bigl(\|Y\|_{\mbb{S}^\infty}^2,
\|(Z^0,Z^1)\|_{\mbb{H}^2_{\rm BMO}}^2\bigr) \\
&\leq \underset{(t,\omega)\in[0,T]\times \Omega}{{\rm ess}\sup}\Bigl(|Y_t|^2+\mbb{E}\Bigl[\int_t^T (|Z_s^0|^2+|Z_s^1|^2)ds|\calf_t^{0,1}\Bigr]\Bigr), \nn
\end{split}
\ee
we obtain
\be
\begin{split}
\|(Z^0,Z^1)\|_{\mbb{H}^2_{\rm BMO}}^2\leq 2\|F\|_{\infty}^2+4c_\gamma^2 \|(z^0,z^1)\|^4_{\mbb{H}^2_{\rm BMO}}. \nn
\end{split}
\ee

We now try to find an $R>0$ such that
\be
2\|F\|_{\infty}^2+4 c_\gamma^2 R^4\leq R^2 \nn
\ee
holds. By completing the square, one sees that this is solvable if and only if  $\|F\|_{\infty}\leq 1/(4\sqrt{2}c_\gamma)$,
and in this case, we can choose $R=2\|F\|_{\infty}$. This proves the first statement.
Moreover, by rearranging the first inequality in $(\ref{BR-middle})$, we get
\be
|Y_t|^2\leq \|F\|_{\infty}^2+\frac{1}{2}\|Y\|_{\mbb{S}^\infty}^2+2c_\gamma^2 \|(z^0,z^1)\|^4_{\mbb{H}^2_{\rm BMO}} \nn
\ee
and hence
\be
\|Y\|_{\mbb{S}^\infty}^2\leq 2\|F\|_{\infty}^2+4c_\gamma^2 \|(z^0,z^1)\|^4_{\mbb{H}^2_{\rm BMO}}. \nn
\ee
This yields $\|Y\|_{\mbb{S}^\infty}^2\leq R^2$ if $(z^0,z^1)\in \calb_R$.
\end{proof}
\end{proposition}

Now, we provide the first main result of this paper.
\begin{theorem}
\label{th-mfg-existence}
Let Assumption~\ref{assumption-agent} be in force. 
If the terminal function $F$ is small enough in the sense that
\be
\|F\|_{\infty}<\frac{1}{48 C_\gamma}, 
\label{terminal-constraint}
\ee
where $C_\gamma$ is a constant used in $(\ref{f-spread})$, then there exists a unique solution $(Y,Z^0,Z^1)$ to the mean-field BSDE $(\ref{mfg-BSDE})$
in the domain
\be
(Z^0,Z^1)\in \calb_R, \quad \|Y\|_{\mbb{S}^\infty}\leq R \nn
\ee
with $R:=2\|F\|_{\infty}$.
\begin{proof}
Note that the requirement $(\ref{terminal-constraint})$ is more stringent than the one used in 
Proposition~\ref{prop-Gamma-stability}. Thus it suffices to prove that the map $\Gamma$ is a strict contraction.

To prove the contraction, let us consider two arbitrary inputs $z:=(z^0,z^1),~\ac{z}:=(\ac{z}^0,\ac{z}^1)\in\calb_R$.
We set
\be
Z:=(Z^0,Z^1):=\Gamma(z), \quad \ac{Z}:=(\ac{Z}^0,\ac{Z}^1):=\Gamma(\ac{z}), \nn
\ee
and $Y, \ac{Y}$ as the $y$-component of the solution of $(\ref{map-Gamma})$ with input $z$ and $\ac{z}$, respectively.
For notational simplicity, we put 
\be
\Del z:=z-\ac{z}, \quad \Del Y:=Y-\ac{Y}, \quad \Del Z:=Z-\ac{Z}.  \nn
\ee
This gives
\be
\Del Y_t=\int_t^T \bigl(f(z_s^0,z_s^1)-f(\ac{z}_s^0,\ac{z}_s^1)\bigr)ds-\int_t^T \Del Z_s^0 dW_s^0-\int_t^T \Del Z_s^1 dW_s^1, \nn
\ee
and hence by \Ito formula, 
\be
|\Del Y_t|^2+\int_t^T(|\Del Z_s^0|^2+|\Del Z_s^1|^2)ds=\int_t^T 2\Del Y_s (f(z_s^0,z_s^1)-f(\ac{z}_s^0,\ac{z}_s^1))ds-\int_t^T 
2\Del Y_s(\Del Z_s^0 dW_s^0+\Del Z_s^1 dW_s^1). \nn
\ee
By taking the conditional expectation, we have, from H\"older inequality,
\be
\begin{split}
&|\Del Y_t|^2+\mbb{E}\Bigl[\int_t^T (|\Del Z_s^0|^2+|\Del Z_s^1|^2)ds|\calf_t^\zo\Bigr]
\leq 2\|\Del Y\|_{\mbb{S}^\infty}\mbb{E}\Bigl[\int_t^T |f(z_s^0,z_s^1)-f(\ac{z}_s^0,\ac{z}_s^1)|ds|\calf_t^\zo\Bigr]\\
&~ \leq\frac{1}{2} \|\Del Y\|_{\mbb{S}^\infty}^2+2\Bigl(\mbb{E}\Bigl[\int_t^T |f(z_s^0,z_s^1)-f(\ac{z}_s^0,\ac{z}_s^1)|ds|\calf_t^\zo\Bigr]\Bigr)^2\\
&~ \leq \frac{1}{2}\|\Del Y\|_{\mbb{S}^\infty}^2+2 C_\gamma^2\Bigl(
\mbb{E}\Bigl[\int_t^T (|z_s^0|+|\ac{z}_s^0|+|z_s^1|+|\ac{z}_s^1|+|\ol{\ex}[z_s^0]|+|\ol{\ex}[\ac{z}_s^0]|)
(|\Del z_s^0|+|\Del z_s^1|+|\ol{\ex}[\Del z_s^0]|)ds|\calf_t^\zo\Bigr]\Bigr)^2  \\
&~\leq \frac{1}{2}\|\Del Y\|_{\mbb{S}^\infty}^2+ 2 C_\gamma^2
\Bigl(\mbb{E}\Bigl[\int_t^T (|z_s^0|+|\ac{z}_s^0|+|z_s^1|+|\ac{z}_s^1|+|\ol{\ex}[z_s^0]|+|\ol{\ex}[\ac{z}_s^0]|)^2ds|\calf_t^\zo\Bigr]\Bigr)\\
&\hspace{30mm} \times \Bigl(\mbb{E}\Bigl[\int_t^T (|\Del z_s^0|+|\Del z_s^1|+|\ol{\ex}[\Del z_s^0]|)^2ds|\calf_t^\zo\Bigr]\Bigr) \\
&~\leq \frac{1}{2}\|\Del Y\|_{\mbb{S}^\infty}^2+2 C_\gamma^2
\Bigl( 6 \mbb{E}\Bigl[\int_t^T (|z_s^0|^2+|\ac{z}_s^0|^2+|z_s^1|^2+|\ac{z}_s^1|^2+|\ol{\ex}[z_s^0]|^2+|\ol{\ex}[\ac{z}_s^0]|^2)ds|\calf_t^\zo\Bigr]\Bigr)\\
&\hspace{30mm} \times \Bigl(3 \mbb{E}\Bigl[\int_t^T (|\Del z_s^0|^2+|\Del z_s^1|^2+|\ol{\ex}[\Del z_s^0]|^2)ds|\calf_t^\zo\Bigr]\Bigr). \nn
\end{split}
\ee
By the same technique used in the proof of Proposition~\ref{prop-Gamma-stability} and $(\ref{bmo-comparison})$, we get
\be
\begin{split}
\|\Del Z\|^2_{\mbb{H}^2_{\rm BMO}}&\leq 4 C_\gamma^2\Bigl(12\sup_{\tau\in\calt^\zo}\Bigr\|
\mbb{E}\Bigl[\int_\tau^T (|z_s|^2+|\ac{z_s}|^2)ds|\calf_\tau^\zo\Bigr]\Bigr\|_{\infty}\Bigr)\Bigl(6\sup_{\tau\in\calt^\zo}\Bigr\|\mbb{E}\Bigl[\int_\tau^T |\Del z_s|^2 ds|\calf_\tau^\zo\Bigr]\Bigr\|_{\infty}\Bigr). \nn
\end{split}
\ee

Since $z,\ac{z}\in \calb_R$, we get
\be
\|\Del Z\|^2_{\mbb{H}^2_{\rm BMO}}\leq 576C_\gamma^2 R^2\|\Del z\|^2_{\mbb{H}^2_{\rm BMO}}.\nn
\ee
Hence the map $\Gamma$ on $\calb_R$ becomes contraction if $R<\frac{1}{24 C_\gamma}$.
Under the choice of $R=2\|F\|_{\infty}$, this is equivalent to
\be
\|F\|_{\infty}<\frac{1}{48 C_\gamma}. \nn
\ee
In this case $\Del Z \rightarrow 0$ in $\mbb{H}^2_{\rm BMO}$ under the repeated application of the map $\Gamma$, and it is also clear that
$\Del Y\rightarrow 0$ in $\mbb{S}^\infty$.
The unique fixed point $Z\in \calb_R$ of the map $\Gamma$ and the associated $Y$ gives a unique solution 
to the mean-field BSDE $(\ref{mfg-BSDE})$ in the domain $Z\in \calb_R$.
\end{proof}
\end{theorem}

\begin{remark}
Recall that there is an invariance of the optimal trading strategy under the transformation given by $(\ref{duality-relation})$.
Therefore, for our purposes to obtain an equilibrium model with exponential utility, 
the constraint on the terminal function $F$ in Theorem~\ref{th-mfg-existence}
is not a direct restriction on the absolute size of liability, but  on the size of deviation from its mean:
\be
\bigl|F-\ex[F|\calf_0^\zo]\bigr|=\bigl|F-\ex[F|\calf_0^1]\bigr|. \nn
\ee
\end{remark}

\subsection{Existence of a solution to the mean-field BSDE with a special terminal structure}
Let us provide one special example where the mean-field BSDE $(\ref{mfg-BSDE})$ has, at least,  one solution 
$(Y,Z^0,Z^1)\in \mbb{S}^\infty\times \mbb{H}^2_{\rm BMO}\times \mbb{H}^2_{\rm BMO}$ even when the terminal function $F$ 
does not satisfy the constraint $(\ref{terminal-constraint})$. As a result, we shall see that Theorem~\ref{th-mfg-existence}
is merely an example of sufficient conditions for the well-posedness of the mean-field BSDE $(\ref{mfg-BSDE})$.

Let us rewrite the mean-field BSDE $(\ref{mfg-BSDE})$ with the rescaled variables:~\footnote{Recall that, in this subsection, we  are working on the convention $(F,\gamma)=(F^1,\gamma^1)$.}
\be
(y,z^0,z^1):=(\gamma Y, \gamma Z^0, \gamma Z^1), \quad G=\gamma F, \nn
\ee
which yields
\be
y_t=G+\int_t^T \Bigl(\wh{\gamma}z_s^{0\parallel}\ol{\ex}\Bigl[\frac{1}{\gamma}z_s^{0\parallel}\Bigr]^\top-\frac{\wh{\gamma}^2}{2}\Bigl|
\ol{\ex}\Bigl[\frac{1}{\gamma}z_s^{0\parallel}\Bigr]\Bigr|^2+\frac{1}{2}(|z_s^{0\perp}|^2+|z_s^1|^2)\Bigr)ds-\int_t^T z_s^0 dW_s^0-\int_t^T z_s^1 dW_s^1,
~t\in[0,T]. 
\label{mfg-BSDE-norm}
\ee
We put the following assumption.
\begin{assumption}
\label{assumption-special}
The rescaled terminal function $G$ has an additive form:
\be
G=G^0+G^1, \nn
\ee
where $G^0$ (resp. $G^1$) is a  bounded $\calf^0_T$ (resp. $\calf^1_T$)-measurable random variable.
\end{assumption}

\begin{remark}
In terms of the original liability $F~(=F^1)$, the above condition is equivalent to assume that $F$ has
the following structure:
\be
F=\frac{1}{\gamma}\wt{F}^0+\wt{F}^1, \nn
\ee
where $\wt{F}^0$ (resp. $\wt{F}^1$) is a bounded $\calf_T^0$ (resp. $\calf_T^1$)-measurable random variable.
In the financial market with distribution of agents as specified by Assumption~\ref{assumption-hetero},
this implies that the part of liability dependent on the common noise are distributed as inversely proportional 
to the agents' risk-averseness parameters $(\gamma^i, i\in \mbb{N})$.
\end{remark}

\begin{theorem}
\label{th-special-existence}
Under Assumptions~\ref{assumption-agent} and \ref{assumption-special},
there is, at least, one solution $(y,z^0,z^1)\in \mbb{S}^\infty\times \mbb{H}^2_{\rm BMO}
\times \mbb{H}^2_{\rm BMO}$ to $(\ref{mfg-BSDE-norm})$, or equivalently  a solution $(Y,Z^0,Z^1)\in \mbb{S}^\infty\times \mbb{H}^2_{\rm BMO}
\times \mbb{H}^2_{\rm BMO}$ to $(\ref{mfg-BSDE})$.
\begin{proof}
Consider the following two BSDEs:
\be
\begin{split}
y_t^0&=G^0+\int_t^T \frac{1}{2}|z_s^0|^2ds-\int_t^T z_s^0 dW_s^0, \quad t\in[0,T], \\
y_t^1&=G^1+\int_t^T \frac{1}{2}|z_s^1|^2 ds-\int_t^T z_s^1 dW_s^1, \quad t\in[0,T]. 
\end{split}
\nn
\ee
It is clear that there exists a unique solution $(y^i, z^i)\in \mbb{S}^\infty\times\mbb{H}^2_{\rm BMO}$ with $i=0,1$
by the standard results on qg-BSDEs in \cite{Kobylanski}. 
Since $(y^0,z^0)$ is $\mbb{F}^0$-adapted and so is $z^{0\parallel}$, we have
\be
\ol{\ex}\Bigl[\frac{1}{\gamma}z_s^{0\parallel}\Bigr]=\ol{\ex}\Bigl[\frac{1}{\gamma}\Bigr]z_s^{0\parallel}=\frac{1}{\wh{\gamma}}z_s^{0\parallel}. \nn
\ee
Therefore, $(y^0,z^0)$ also solves the BSDE
\be
y_t^0=G^0+\int_t^T\Bigl(\wh{\gamma}z^{0\parallel}_s\ol{\ex}\Bigl[\frac{1}{\gamma}z_s^{0\parallel}\Bigr]^\top-\frac{\wh{\gamma}^2}{2}
\Bigr|\ol{\ex}\Bigl[\frac{1}{\gamma}z_s^{0\parallel}\Bigr]\Bigr|^2+\frac{1}{2}|z_s^{0\perp}|^2\Bigr)ds-\int_t^T z_s^0 dW_s^0, ~t\in[0,T]. \nn
\ee
It is now clear that $(y,z^0,z^1):=(y^0+y^1, z^0,z^1)$ provides a solution to $(\ref{mfg-BSDE-norm})$. 
\end{proof}
\end{theorem}

\begin{remark}
Under the assumptions used in Theorem~\ref{th-special-existence}, we actually have a closed form solution,
\be
y_t^j=\ln\bigl(\mbb{E}\bigl[\exp(G^j)|\calf_t^j\bigr]\bigr), ~j=0,1. \nn
\ee
This result can be easily confirmed by the Cole-Hopf transformation, $\exp({y_t^j})$.

\end{remark}
\section{Market clearing in the large population limit}
\label{sec-market-clearing}
Finally, in this section, we shall show that the process $(\theta^{\rm mfg}_t,t\in[0,T])\in \mbb{H}^2_{\rm BMO}$ 
defined by $(\ref{theta-mfg})$
in terms of the solution to the mean-field BSDE is actually 
a good approximate of the risk-premium process in the market-clearing equilibrium.
\\

In order to treat the large population limit, we first enlarge the complete probability space $(\Omega,\calf,\mbb{P})$ 
with filtration $\mbb{F}:=(\calf_t)_{t\geq 0}$ as follows.
\bi
\item $\Omega:=\Omega^0\times \prod_{i=1}^\infty \Omega^i$
and $(\calf,\mbb{P})$ is the completion of $\bigl(\calf^0\otimes \bigotimes_{i=1}^\infty \calf^i ,
\mbb{P}^0\otimes \bigotimes_{i=1}^\infty \mbb{P}^i\bigr)$. $\mbb{F}$ denotes the complete and the right-continuous augmentation of $(\calf^0_t\otimes \bigotimes_{i=1}^\infty \calf_t^i)_{t\geq 0}$.
$\mbb{E}[\cdot]$ denotes the expectation with respect to $\mbb{P}$.~\footnote{
For general results on the construction of a (countable or even uncountable) product probability space, 
see, for example,  Klenke~\cite{Klenke}[Section~14.3].} 
\ei

Suppose that the financial market is defined as in Assumption~\ref{assumption-market} with the process $\mu$ given by 
$(\mu_t:=\sigma_t\theta_t^{\rm mfg}, t\in[0,T])$.
Notice that the process $\theta^{\rm mfg}$ (and hence also $\mu$) is $\mbb{F}^0$-adapted and consistent with our assumption on the information structure. In particular, this means that each agent (agent-$i$) can implement her strategy based on the common 
and her own idiosyncratic informations $\mbb{F}^0\otimes \mbb{F}^i$ without taking care of idiosyncratic noise of the other agents.
Therefore, for each $i\in \mbb{N}$, the optimal trading strategy of the agent-$i$ is provided by, as in $(\ref{p-i-optimal})$, 
\be
p_t^{i,*}~(=(\pi_t^{i,*})^\top\sigma_t)=Z_t^{i,0\parallel}-\frac{\wh{\gamma}}{\gamma^i}\ol{\ex}[\calz_t^{0\parallel}], \quad t\in[0,T].  \nn
\ee
Here,  $Z^{i,0}$ is associated to the solution of the BSDE (see, $(\ref{BSDE-agent-i})$)
\be
\begin{split}
Y_t^i&=F^i+\int_t^T \Bigl(\wh{\gamma}Z_s^{i,0\parallel}\ol{\ex}[\calz_s^{0\parallel}]^\top-\frac{\wh{\gamma}^2}{2\gamma^i}|\ol{\ex}[\calz_s^{0\parallel}]|^2+\frac{\gamma^i}{2}(|Z_s^{i,0\perp}|^2+|Z_s^{i}|^2)\Bigr)ds \\
&\quad -\int_t^T Z_s^{i,0}dW_s^0-\int_t^T Z_s^i dW_s^i, ~ t\in[0,T], 
\end{split}
\label{mfg-agent-i}
\ee
and $\calz^0$  is associated to the solution of
the mean-field BSDE $(\ref{mfg-BSDE-org})$~\footnote{For a clear distinction, we have changed the symbols.}
\be
\caly_t^1=F^1+\int_t^T \Bigl(\wh{\gamma}\calz_s^{0\parallel}\ol{\ex}[\calz_s^{0\parallel}]^\top
-\frac{\wh{\gamma}^2}{2\gamma^1}|\ol{\ex}[\calz_s^{0\parallel}]|^2+\frac{\gamma^1}{2}(|\calz_s^{0\perp}|^2+
|\calz_s^1|^2)\Bigr)ds-\int_t^T \calz_s^0 dW_s^0-\int_t^T \calz_s^1 dW_s^1,~ t\in[0,T]. \nn
\ee
From Theorem~\ref{th-mfg-existence} and Theorem~\ref{th-special-existence}, we already know that there 
exists a bounded solution (possibly not unique) $(\caly^1,\calz^0,\calz^1)\in \mbb{S}^\infty\times \mbb{H}^2_{\rm BMO}
\times \mbb{H}^2_{\rm BMO}$ to the mean-field BSDE $(\ref{mfg-BSDE-org})$ under certain conditions.

Here is the last main result of this paper.
\begin{theorem}
\label{th-mfg-clearing}
Let Assumptions~\ref{assumption-market} and \ref{assumption-hetero} be in force.
Assume in addition that there is a bounded solution $(\caly^1,\calz^0,\calz^1)\in \mbb{S}^\infty\times \mbb{H}^2_{\rm BMO}
\times \mbb{H}^2_{\rm BMO}$  to the mean-field BSDE $(\ref{mfg-BSDE-org})$ and that
we select an arbitrary but fixed solution from it to define the risk-premium process $(\theta_t^{\rm mfg}:=-\wh{\gamma}\ol{\ex}[\calz_t^{0\parallel}]^\top,~t\in[0,T])$.
Then,  $\theta^{\rm mfg}$ clears the market in the large population limit in the sense that, 
the agents' optimal trading strategies $(\pi^{i,*})_{i\in \mbb{N}}$ satisfy the estimate
\be
\mbb{E}\int_0^T\Bigr|\frac{1}{N}\sum_{i=1}^N \pi_t^{i,*}\Bigr|^2 dt\leq \frac{C}{N} \nn
\ee
with some constant $C>0$ uniformly in $N\in \mbb{N}$.
\begin{proof}
For a given $\theta^{\rm mfg}~(=-\wh{\gamma}\ol{\ex}[\calz^{0\parallel}]^\top)$, which is in $\mbb{H}^2_{\rm BMO}$ by $(\ref{bmo-comparison})$, 
it follows from Corollary~\ref{corollary-existence} that the BSDE $(\ref{mfg-agent-i})$ has a unique bounded solution 
$(Y^i,Z^{i,0},Z^i)\in \mbb{S}^\infty\times \mbb{H}^2_{\rm BMO}\times \mbb{H}^2_{\rm BMO}$
for every $i\in \mbb{N}$.
In particular, the uniqueness of the solution
implies  $(Y^1,Z^{1,0},Z^1)=(\caly^1,\calz^0,\calz^1)$, the latter of which is the one used to define the process $\theta^{\rm mfg}$.
Thus we have $\ol{\ex}[\calz^{0\parallel}]=\ol{\ex}[Z^{1,0\parallel}]$.

The uniqueness of the solution of $(\ref{mfg-agent-i})$ also implies, by Yamada-Watanabe Theorem~(see, for example, \cite{Carmona-Delarue-2}[Theorem 1.33]),  that there exists a some measurable function $\Phi$ such that
\be
(Y^i, Z^{i,0},Z^i)=\Phi\Bigl(W^0, \xi^i,\gamma^i,W^i,(\theta^{\rm mfg},F^i)\Bigr), ~\forall i\in\mbb{N} \nn
\ee
where $\Phi$ depends only on the joint distribution $\call\bigl(W^0, \xi^i,\gamma^i,W^i,(\theta^{\rm mfg},F^i)\bigr)$.
Since $\theta^{\rm mfg}$ is $\mbb{F}^0$-adapted and $F^i$ is $\calf^0$-conditionally i.i.d, 
this expression implies that the solutions $\{(Y^i, Z^{i,0},Z^i),~i\in\mbb{N}\}$ are $\calf^0$-conditionally i.i.d.

Since $\pi^{i,*}_t=(\sigma_t\sigma_t^\top)^{-1}\sigma_t (p^{i,*}_t)^\top$, we have,  from Assumption~\ref{assumption-market} (ii),
\be
\mbb{E}\int_0^T\Bigr|\frac{1}{N}\sum_{i=1}^N \pi_t^{i,*}\Bigr|^2 dt\leq C\mbb{E}\int_0^T\Bigr|\frac{1}{N}\sum_{i=1}^N p_t^{i,*}\Bigr|^2 dt \nn
\ee
with some constant $C$. Since $\calz^0=Z^{1,0}$, we have
\be
\begin{split}
\frac{1}{N}\sum_{i=1}^N p_t^{i,*}=\frac{1}{N}\sum_{i=1}^N(Z_t^{i,0\parallel}-\ol{\ex}[Z_t^{1,0\parallel}])+\frac{1}{N}\sum_{i=1}^N
\Bigl(1-\frac{\wh{\gamma}}{\gamma^i}\Bigr)\ol{\ex}[Z_t^{1,0\parallel}], \nn
\end{split}
\ee
which then yields 
\be
\begin{split}
\mbb{E}\int_0^T \Bigr|\frac{1}{N}\sum_{i=1}^N p_t^{i,*}\Bigr|^2 dt & \leq 2\mbb{E}\int_0^T \Bigl| \frac{1}{N}\sum_{i=1}^N(Z_t^{i,0\parallel}-\ol{\ex}[Z_t^{1,0\parallel}])\Bigr|^2dt+
2\wh{\gamma}^2 \mbb{E}\int_0^T\Bigl|\frac{1}{N}\sum_{i=1}^N\Bigl(\frac{1}{\wh{\gamma}}-\frac{1}{\gamma^i}\Bigr)
\ol{\ex}[Z_t^{1,0\parallel}]\Bigr|^2 dt \\
&\hspace{-10mm}=2\mbb{E}\int_0^T \Bigl| \frac{1}{N}\sum_{i=1}^N(Z_t^{i,0\parallel}-\ol{\ex}[Z_t^{1,0\parallel}])\Bigr|^2dt+
2\wh{\gamma}^2 \mbb{E}\Bigl[\Bigl|\frac{1}{N}\sum_{i=1}^N\Bigl(\frac{1}{\wh{\gamma}}-\frac{1}{\gamma^i}\Bigr)\Bigr|^2\Bigr]
\mbb{E}\int_0^T |\ol{\ex}[Z_t^{1,0\parallel}]|^2 dt, \nn
\end{split}
\ee
where we used the independence of $(\gamma^i)_{i\in \mbb{N}}$ and $\calf^0$ in the second line.
Since $(1/\gamma^i)_{i=1}^N$ are i.i.d. and $(Z_t^{i,0\parallel})_{i=1}^N$ are $\calf^0$-conditionally i.i.d., 
the cross terms of the two expectations both vanish. Hence, we obtain
\be
\begin{split}
\mbb{E}\int_0^T \Bigr|\frac{1}{N}\sum_{i=1}^N p_t^{i,*}\Bigr|^2 dt &\leq \frac{2}{N^2}\sum_{i=1}^N\mbb{E}\int_0^T |Z_t^{0,i}-\ol{\ex}[Z_t^{0,1}]|^2dt+\frac{2\wh{\gamma}^2}{N^2}
\sum_{i=1}^N\mbb{E}\Bigl[\Bigr|\frac{1}{\wh{\gamma}}-\frac{1}{\gamma^i}\Bigr|^2\Bigr]\mbb{E}\int_0^T |Z_t^{1,0}|^2 dt\\
&\leq \frac{4}{N}\Bigl(1+\frac{\wh{\gamma}^2}{\ul{\gamma}^2}\Bigr)\|Z^{0,1}\|^2_{\mbb{H}^2_{\rm BMO}}, \nn
\end{split}
\ee
where we have used Jensen's inequality and $(\ref{ineq-energy})$ with $n=1$. This gives the desired result.
\end{proof}
\end{theorem}

Combined with Theorems~\ref{th-mfg-existence} and \ref{th-special-existence}, the result of Theorem~\ref{th-mfg-clearing} implies that
$\theta^{\rm mfg}$ given by $(\ref{theta-mfg})$ achieves the market-clearing condition in the large-$N$ limit in the sense that
\be
\lim_{N\rightarrow \infty} \frac{1}{N}\sum_{i=1}^N \pi_t^{i,*}=0, ~\text{$\mbb{P}\otimes dt$-a.e.} \nn
\ee
at least if, for every $i$,  $|F^i-\ex[F^i|\calf^i_0]|$ is small enough or $F^i$ has an additive form.
It mean, economically, the excessive demand/supply per capita converges to zero $\mbb{P}\otimes dt$-a.e. in the large population limit.
To fully understand the general conditions for the existence of mean-field equilibrium, we need further study 
on the mean-field BSDE $(\ref{mfg-BSDE})$. 

\begin{remark}
Since the contraction mapping approach by Tevzadze~\cite{Tevzadze}[Proposition 1] used in our proof of Theorem~\ref{th-mfg-existence}
is applicable to multi-dimensional setups, it may be used to prove the existence of  the  equilibrium 
among the finite number of  (say, $N$) agents, by directly solving the coupled system of qg-BSDEs $(\ref{qg-BSDE-system})$ (with appropriate 
modifications of equations as well as the admissible spaces). 
However, this approach requires the assumption of  perfect information as well as the smallness of the {\bf{total}} size of liabilities among the agents; 
$\|\sqrt{\sum_{i=1}^N |F^i|^2}\|_{\infty}$. This means that the constraint on the liabilities becomes more and more stringent as the population grows.
\end{remark}

\begin{remark}
Unfortunately, uniqueness of the market equilibrium is not known. We expect that 
we may have uniqueness when the conditions for Theorem 4.1 are satisfied, i.e. small 
enough $\|F\|_{\infty}$ for which the mean-field BSDE $(\ref{mfg-BSDE})$ has a unique solution in a specified domain. 
However, in this paper, we have just tried ``guess-and-check" strategy to find a candidate of equilibrium.  Therefore, although it seems unlikely, we cannot exclude the existence of a market-clearing equilibrium that is not characterized by our BSDE  $(\ref{mfg-BSDE})$. 
\end{remark}

\section{Conclusion and discussions}
\label{sec-conclusion}
In this paper, we studied a problem of equilibrium price formation among many investors with  exponential utility.
We allowed the agents to be heterogeneous in their initial wealth, risk-averseness parameter, as well as stochastic liability at the terminal time.
We showed that the equilibrium risk-premium process of risky stocks 
is characterized by the solution to  a novel mean-field BSDE,  whose driver has quadratic growth both in the stochastic integrands 
and in their conditional expectations. We proved the existence of a solution to the mean-field BSDE under several conditions
and showed that the resultant risk-premium process actually clears the market in the large population limit.

Let us point out several directions for further research. Extending the proposed technique to 
the price formation with consumptions or different interactions among the agents is an important research topic.
Applications to macroeconomic models in the presence of production may also be possible. 
Finally, the novel mean-field BSDE $(\ref{mfg-BSDE-org})$ deserves further investigations to relax its
well-posedness conditions. The same type of BSDEs are expected to appear in various applications of 
MFGs and the martingale method~\cite{Hu-Imkeller}.

\subsubsection*{Acknowledgments}
 M.S. is supported by JSPS KAKENHI Grant Number 23KJ0648.

\appendix
\section{Appendix}

\subsection{Proof of Lemma~\ref{lemma-Z-norm}}
\label{sec-A-1}
By \Ito formula, we have,
\be
\begin{split}
de^{2y_t}&=e^{2 y_t}\bigl(2z_t^{0\parallel}\theta_t+ |\theta_t|^2-\bigl(|z_t^{0\perp}|^2+|z_t^1|^2)+2
(|z_t^0|^2+|z_t^1|^2)\bigr)dt\\
&\quad+2 e^{2y_t}(z_t^0 dW_t^0+z_t^1 dW_t^1)\\
&\geq e^{2y_t}\bigl(-|z_t^{0\parallel}|^2-\bigl(|z_t^{0\perp}|^2+|z_t^1|^2)+2(|z_t^0|^2+|z_t^1|^2)\bigr)dt\\
&\quad+2 e^{2 y_t}(z_t^0 dW_t^0+z_t^1 dW_t^1)\\
&\geq e^{2y_t}(|z_t^0|^2+|z_t^1|^2)dt+2 e^{2y_t}(z_t^0dW_t^0+z_t^1 dW_t^1)
\end{split}
\nn
\ee
and thus, for any $t\in[0,T]$,
\be
e^{2 y_T}-e^{2y_t}\geq \int_t^T e^{2 y_s}(|z_s^0|^2+|z_s^1|^2)ds+\int_t^T 2 e^{2 y_s}(z_s^0dW_s^0+z_s^1 dW_s^1). \nn
\ee
Thus, it is easy to obtain
$
\|z\|^2_{\mbb{H}^2_{\rm BMO}}:=\sup_{\tau\in\calt^\zo}\Bigl\|\mbb{E}\Bigl[\int_\tau^T (|z^0_s|^2+|z_s^1|^2)ds|\calf_\tau^\zo\Bigr]\Bigr\|_{\infty}
\leq e^{4 \|y\|_{\mbb{S}^\infty}}. \qquad\quad \square
$

\subsection{Proof of Theorem~\ref{th-qgBSDE-existence}~(Continued)}
\label{sec-A-2}
Let us define a smooth convex function $\phi:\mbb{R}\rightarrow \mbb{R}_+$ satisfying
\be
\phi(0)=0, \quad \phi^\prime(0)=0, \nn
\ee 
whose concrete form is to be determined later.
We consider $m, n\in \mbb{N}$ such that $m\geq n$ and put
\be
\Del y^{n,m}:=y^n-y^m, \quad \Del z^{n,m;0}:=z^{n,0}-z^{m,0}, \quad \Del z^{n,m;1}:=z^{n,1}-z^{m,1}. \nn
\ee
Note that $\Del y^{n,m}\geq 0$ since $m\geq n$. From \Ito formula, we obtain for any $t\in[0,T]$,
\be
\begin{split}
&\phi(\Del y_t^{n,m})+\int_t^T \frac{1}{2}\phi^{\prime\prime}(\Del y_s^{n,m})(|\Del z_s^{n,m;0}|^2+|\Del z_s^{n,m;1}|^2)ds\\
&=\int_t^T\phi^\prime(\Del y^{n,m}_s)\bigl[-\frac{1}{2}(|\theta_s|^2\wedge n)+\frac{1}{2}(|z_s^{n,0\perp}|^2+|z^{n,1}_s|^2)+\frac{1}{2}(|\theta_s|^2\wedge m)-\frac{1}{2}(|z_s^{m,0\perp}|^2+|z_s^{m,1}|^2)\bigr]ds\\
&\quad-\int_t^T \phi^\prime(\Del y_s^{n,m})(\Del z_s^{n,m;0}d\wt{W}_s^0+\Del z_s^{n,m;1}d\wt{W}_s^1).  \nn
\end{split}
\ee
Since $\phi(y), \phi^\prime(y)\geq 0, \forall y\geq 0$, we get
\be
\begin{split}
&\mbb{E}\int_0^T \frac{1}{2}\phi^{\prime\prime}(\Del y^{n,m}_s)(|\Del z_s^{n,m;0}|^2+|\Del z_s^{n,m;1}|^2)ds\\
&\quad \leq \mbb{E}\int_0^T \frac{1}{2}\phi^\prime(\Del y_s^{n,m})\bigl(|\theta_s|^2+|z_s^{n,0}|^2+|z_s^{n,1}|^2\bigr)ds\\
&\quad \leq \mbb{E}\int_0^T \phi^\prime(\Del y_s^{n,m})\bigl(|\theta_s|^2+|z_s^{n,0}-z_s^0|^2+|z_s^{n,1}-z_s^1|^2+|z_s^0|^2+|z_s^1|^2\bigr)ds. 
\end{split}
\label{phi-prpr}
\ee
We now choose the function $\phi$ as 
\be
\phi(y):=\frac{1}{8}[e^{4y}-4y-1], \nn
\ee
which gives $\phi^\prime(y)=\frac{1}{2}[e^{4y}-1]$ and $\phi^{\prime\prime}(y)=2e^{4y}$. In particular, 
we have $\phi^{\prime\prime}(y)=4\phi^\prime(y)+2$. This yields, from $(\ref{phi-prpr})$,
\be
\begin{split}
&\mbb{E}\int_0^T[2\phi^\prime(\Del y_s^{n,m})+1](|\Del z^{n,m;0}_s|^2+|\Del z_s^{n,m;1}|^2)ds\\
&\quad\leq \mbb{E}\int_0^T \phi^\prime(\Del y^{n,m}_s) (|\theta_s|^2+|z_s^{n,0}-z_s^0|^2+|z_s^{n,1}-z_s^1|^2+|z_s^0|^2+|z_s^1|^2)ds.
\end{split}
\label{phi-pr}
\ee

Note that, since $(\Del y^{n,m})_{m\geq n}$ are bounded and strongly convergent $\Del y^{n,m}\rightarrow y^n-y$ as $m\rightarrow \infty$,
we also have, under an appropriate subsequence (still denoted by $m$), the following weak convergence in $\mbb{H}^2$;
\be
\sqrt{2\phi^\prime(\Del y^{n,m})+1}\Del z^{n,m;j}\rightharpoonup \sqrt{2\phi^\prime(y^n-y)+1}(z^{n,j}-z^j), \quad {\text{as $m\rightarrow \infty$}}, \nn
\ee
with $j=0,1$. Hence, by \cite{Brezis}[Proposition 3.5] and monotone convergence, we obtain from $(\ref{phi-pr})$,
\be
\begin{split}
&\mbb{E}\int_0^T [2\phi^\prime(y^n_s-y_s)+1](|z_s^{n,0}-z_s^0|^2+|z_s^{n,1}-z_s^1|^2)ds\\
&\quad \leq \liminf_{m\rightarrow \infty}\mbb{E}\int_0^T [2\phi^\prime(\Del y_s^{n,m})+1](|\Del z_s^{n,m;0}|^2+|\Del z_s^{n,m;1}|^2)ds\\
&\quad \leq \mbb{E}\int_0^T\phi^\prime(y^n_s-y_s)(|\theta_s|^2+|z_s^{n,0}-z_s^0|^2+|z_s^{n,1}-z_s^1|^2+|z_s^0|^2+|z_s^1|^2)ds. \nn
\end{split}
\ee
By rearranging the $|z^{n,j}-z^j|^2$-terms $(j=0,1)$, we get
\be
\begin{split}
&\mbb{E}\int_0^T [\phi^\prime(y^n_s-y_s)+1](|z_s^{n,0}-z_s^0|^2+|z_s^{n,1}-z_s^1|^2)ds\\
&\quad \leq  \mbb{E}\int_0^T\phi^\prime(y^n_s-y_s)(|\theta_s|^2+|z_s^0|^2+|z_s^1|^2)ds. \nn
\end{split}
\ee
Since the right-hand side converges to zero as $n\rightarrow \infty$ by the  monotone convergence theorem, we obtain
\be
z^{n,0}\rightarrow z^0, \quad z^{n,1}\rightarrow z^1, \quad {\text{strongly in $\mbb{H}^2$. }} \nn
\ee
Then, from Burkholder-Davis-Gundy (BDG) inequality~\footnote{See, for example, \cite{Protter}[Thorem 48 in IV]. }, 
it implies that the following convergence holds for $(j=0,1)$ under an appropriate subsequence,   
\be
\sup_{t\in[0,T]}\Bigr|\int_t^T(z_s^{n,j}-z_s^j)d\wt{W}_s^j\Bigr|\rightarrow 0, \quad {\text{$\mbb{Q}$-a.s. as $n\rightarrow \infty$}} \nn
\ee
so is the case for $\sup_{t\in[0,T]}|y^n_t-y_t|$. It is now easy to see $(y,z^0,z^1)\in \mbb{S}^\infty\times \mbb{H}^2_{\rm BMO}\times 
\mbb{H}^2_{\rm BMO}$ actually solves the BSDE $(\ref{BSDE-norm-Q})$.
The uniqueness of the solution follows exactly in the same way as in Theorem~\ref{th-uniqueness}. $\qquad\square$

\subsection{Proof of Lemma~\ref{lemma-mfg-bmo}}
\label{sec-A-3}
By applying \Ito formula to $e^{2\gamma Y_t}$ and using $|Z_t^0|^2=|Z_t^{0\parallel}|^2+|Z_t^{0\perp}|^2$, we get
\be
\begin{split}
d(e^{2\gamma Y_t})&=e^{2\gamma Y_t}\bigl(2\gamma dY_t+2\gamma^2 (|Z_t^0|^2+|Z_t^1|^2)dt\bigr)\\
&=e^{2\gamma Y_t}\bigl(-2\gamma\wh{\gamma}Z_t^{0\parallel}\ol{\ex}[Z_t^{0\parallel}]^\top+\wh{\gamma}^2|\ol{\ex}[Z_t^{0\parallel}]|^2
-\gamma^2(|Z_t^{0\perp}|^2+|Z_t^1|^2)+2\gamma^2(|Z_t^0|^2+|Z_t^1|^2)\bigr)dt\\
&\quad+e^{2\gamma Y_t}2\gamma (Z_t^0dW_t^0+Z_t^1 dW_t^1)\\
&\geq e^{2\gamma Y_t}\gamma^2(|Z_t^0|^2+|Z_t^1|^2)dt+e^{2\gamma Y_t}2\gamma(Z_t^0 dW_t^0+Z_t^1 dW_t^1).  \nn
\end{split}
\ee
Hence, for any $\tau \in \calt^{1,0}$, 
\be
\mbb{E}\bigl[e^{2\gamma Y_T}|\calf_\tau^{1,0}\bigr]\geq \mbb{E}\bigl[e^{2\gamma Y_T}-e^{2\gamma Y_\tau}|\calf_\tau^{1,0}\bigr]\geq \mbb{E}\Bigl[\int_\tau^T e^{2\gamma Y_s}\gamma^2 (|Z_s^0|^2+
|Z_s^1|^2)ds|\calf_\tau^{1,0}\Bigr], \nn 
\ee
which then gives~
$
\displaystyle
\|(Z^0,Z^1)\|_{\mbb{H}^2_{\rm BMO}}^2\leq \frac{1}{\ul{\gamma}^2}\exp\bigl({4\ol{\gamma}\|Y\|_{\mbb{S}^\infty}}\bigr). 
\qquad\quad \square
$



\end{document}